\documentclass[twocolumn, resetfootnote]{aastex7}
\graphicspath{{./}{../pictures/}}
\received{}
\revised{}
\accepted{}

\shorttitle{Quasi-stars}
\shortauthors{Hassan et al.}

\usepackage{graphicx}
\usepackage{amsmath}	
\usepackage{amssymb}	
\usepackage{cases}
\usepackage{array,multirow}
\usepackage{upgreek}
\usepackage{textcomp}
\usepackage{lineno}
\usepackage{subcaption}
\usepackage{mhchem}
\usepackage{longtable}

\usepackage{hyperref}
\usepackage[referable]{threeparttablex}
\usepackage{booktabs, dcolumn}

\def\apjl{ApJL}

\def\aap{A\&A}

\def\apjs{ApJ Supp.}

\usepackage[normalem]{ulem}

\begin{document}

\title{The Growth of the Central Black Holes in  Quasi-stars}

\correspondingauthor{Jake B. Hassan}
\email{jakebhassan@gmail.com}
\author[0009-0004-5685-6155]{Jake B. Hassan}
\affil{Department of Physics and Astronomy, Stony Brook University, Stony Brook, NY 11794-3800, USA}
\email{jakebhassan@gmail.com}
\author[0000-0002-3635-5677]{Rosalba Perna}
\affil{Department of Physics and Astronomy, Stony Brook University, Stony Brook, NY 11794-3800, USA}
\email{rosalba.perna@stonybrook.edu}

\author[0000-0002-8171-8596]{Matteo Cantiello}
\affil{Center for Computational Astrophysics, Flatiron Institute, 162 5th Avenue, New York, NY 10010, USA}
\affil{Department of Astrophysical Sciences, Princeton University, Princeton, NJ 08544, USA}
\email{mcantiello@flatironinstitute.org}
\author[0000-0001-5032-1396]{Philip J. Armitage}
\affil{Department of Physics and Astronomy, Stony Brook University, Stony Brook, NY 11794-3800, USA}
\affil{Center for Computational Astrophysics, Flatiron Institute, 162 5th Avenue, New York, NY 10010, USA}
\email{philip.armitage@stonybrook.edu}
\author[0000-0003-0936-8488]{Mitchell C. Begelman}
\affiliation{JILA, University of Colorado and National Institute of Standards and Technology, 440 UCB, Boulder, CO 80309-0440, USA}
\affiliation{Department of Astrophysical and Planetary Sciences, University of Colorado, 391 UCB, Boulder, CO 80309-0391, USA}
\email{mitch@jila.colorado.edu}

\author[0000-0003-2012-5217]{Taeho Ryu}
\affiliation{JILA, University of Colorado and National Institute of Standards and Technology, 440 UCB, Boulder, CO 80309-0440, USA}
\affiliation{Department of Astrophysical and Planetary Sciences, University of Colorado, 391 UCB, Boulder, CO 80309-0391, USA}
\email{taeho.ryu@colorado.edu}
\affiliation{Max-Planck-Institut für Astrophysik, Karl-Schwarzschild-Straße 1, 85748 Garching bei München, Germany}

\begin{abstract}
Observations by JWST have confirmed
the presence of supermassive black holes (BHs) at redshifts $z\gtrsim10$, lending support to scenarios in which BHs experience rapid growth through intense gas accretion. Here we investigate the growth of a BH embedded at the center of a quasi-star, a theoretically predicted object formed via direct collapse.  In a quasi-star, the central BH accretes at a highly super-Eddington rate, while the excess energy is transported outward by convection and radiated at approximately the Eddington luminosity of the entire star.  We employ the open-source stellar evolution code \texttt{MESA} to construct quasi-star models and follow the time-dependent growth of the central BH under different prescriptions for the accretion rate at the inner boundary $R_i$, and further considering
the effect of winds. For the case $R_i=NR_{\rm B}$, where 
$N$ is a constant and
$R_{\rm B}$ is the Bondi radius corresponding to the mass of the BH and the gas infalling onto it, our models terminate when the BH mass reaches a critical value $M_{\mathrm{crit}}(N)=c_{s,i}^3/(12\sqrt{N^3G^3\pi\rho_i})$ (where $c_{s,i}$ and $\rho_i$ are the sound speed and density at $R_i$, respectively), a limit we also derive analytically. Models that feature an inner convective region matched to an outer adiabatic envelope exhibit BH growth up to approximately  $M_{\mathrm{BH}}/M_\star\simeq 0.33$, largely independent of the stellar mass $M_\star$ itself. This ratio is approximately preserved even in the presence of mass loss, as several properties of the model are independent of the quasi-star's total mass.
\end{abstract}


\section{Introduction} 
\label{sec:intro}

One of the central challenges in modern astrophysics is explaining the
rapid emergence of supermassive black holes (SMBHs) within the first
billion years of the universe. Observations of quasars at redshifts $z
\gtrsim 6$ imply that black holes with masses $\gtrsim
10^9\, M_\odot$ were already in place less than a Gyr after the Big Bang \citep{Fan2006,Banados2018}. The conventional growth of black holes (BHs)
from stellar-mass seeds through Eddington-limited accretion struggles
to meet these stringent timescales, prompting the exploration of more
rapid or exotic formation pathways. Among these, quasi-stars
—massive, extended, radiation-pressure supported envelopes
surrounding growing BH seeds—have long been proposed as a
viable mechanism for early black hole growth
\citep{Begelman2006,Begelman2008,Volonteri2010,Ball2011}.

Quasi-stars form when a dense gas cloud undergoes direct collapse of its core region,
leading to the formation of a central BH embedded within a massive
envelope. The envelope feeds the BH through a convective,
radiation-dominated accretion flow, while carrying away the liberated energy and ultimately radiating most of it away at a star-like photosphere.  Because the
quasi-star envelope absorbs and reprocesses the accretion luminosity,
the BH can grow at rates well above the Eddington limit for isolated
accretion, reaching intermediate masses ($\sim 10^3$--$10^4\,
M_\odot$) or higher on short timescales of less than a few Myr
\citep{Begelman2008}. Models of quasi-stars must make choices for several physically uncertain processes, including the efficiency of energy transport in the envelope, the appropriate inner boundary conditions, the treatment of density inversions, and the strength of winds \citep{Dotan2011,Fiacconi2017}. These processes determine the quasi-stars' hydrostatic stability, and, given some assumed ongoing accretion rate, the mass of the BH that emerges at the end of the quasi-star phase. Notably, \citet{Ball2011,Ball2012}, constructing quasi-star models with the Cambridge \texttt{STARS} evolution code and a free-fall model for the inner region \citep{Begelman2008} found that the central BH cannot grow
beyond a few percent of the total quasi-star mass $M_\star$. More recently, however, \citet{Coughlin2024} revisited the internal structure of
quasi-stars using a more physically realistic treatment of the inner
boundary condition at the BH's radius of influence. Their
results demonstrated that a substantially larger fraction of the
envelope mass can be accreted, leading to
a BH mass $\sim 0.62 M_\star$, and significantly
enhancing the plausibility of quasi-stars as progenitors of the SMBHs
powering high-redshift quasars.

At the same time, recent observational discoveries have revived
interest in the quasi-star hypothesis. Deep imaging from the
\textit{James Webb Space Telescope} (JWST) has revealed a population
of faint, red, compact sources—dubbed ``Little Red Dots'' (LRDs)— at
redshifts $z \sim 6$--$10$.
Their spectral energy distribution is characterized by a ``V"-shaped spectrum in the rest-frame continuum between the UV and the optical, with a turnover near the Balmer break
(e.g. \citealt{Labbe2023,Furtak2023,Barro2024,Greene2024,Kokorev2024,Matthee2024,Setton2024,Akins2025,Koceski2025,Taylor2025}). Most of them display broad, typically exponentially-shaped lines and
lack X-ray emission, possibly an indication of high column densities.  

The physical origin
of LRDs remains uncertain, with possibilities ranging from heavily
dust-enshrouded star-forming galaxies to black
holes accreting via geometrically thick super-Eddington flows in an early growth phase \citep{Liu2025}. Quasi-stars naturally predict red,
compact, low-temperature emission due to their extended envelopes and
reprocessed luminosity, and have therefore emerged as compelling
candidates to explain at least a subset of the LRD population \citep{Begelman2025}. 

In this work, we investigate the connection between quasi-stars and the rapid growth of massive BH seeds by constructing detailed models of quasi-star structure and evolution, motivated by the recent theoretical and observational developments discussed above. We employ the stellar evolution code \texttt{MESA} \citep{paxton2011,paxton2013,paxton2015,paxton2018,paxton2019} to model the time-dependent evolution of quasi-stars and the corresponding growth of their central BHs. Our analysis focuses on assessing the impact of the adopted inner boundary condition, as well as of mass loss, on the final BH mass, with the goal of determining whether the central BH can ultimately grow to a significant fraction of the total quasi-star mass.

Our paper is organized as follows. In Section~\ref{sec:numMethods}, we describe the numerical methods employed in this work. In Section~\ref{sec:BallModel}, we present our implementation of the framework developed by \citet{Ball2011,Ball2012}, and derive an analytical expression for the maximum BH mass attainable within their model, demonstrating excellent agreement with the numerical evolution obtained using \texttt{MESA}. In Section~\ref{sec:CoughlinModel}, we detail our implementation of the inner boundary conditions following \citet{Coughlin2024}, and inclusive of the more recent updates by \citet{Begelman2025}, while
Section~\ref{sec:windstheory} describes our treatment of mass loss.
The results of the numerical evolution within this framework are presented in Section~\ref{sec:results}, studying first models without winds (Section~\ref{sec:noWinds}) and then considering the effect of winds (Section~\ref{sec:winds}). Last, our conclusions and summary are provided in Section~\ref{sec:summary}.

\section{Structure and evolution of quasistars}

\subsection{Numerical methods}
\label{sec:numMethods}

We use version r24.08.1 of the open-source stellar evolution code Modules for Experiments in Stellar Astrophysics (\texttt{MESA}; \citealt{paxton2011,paxton2013,paxton2015,paxton2018,paxton2019}) to construct evolutionary models of quasi-stars. \texttt{MESA} is a one-dimensional numerical tool designed to model stellar structure and evolution across a broad range of stellar masses, chemical compositions, and evolutionary stages. It solves the coupled stellar structure equations using an implicit scheme with adaptive timesteps and incorporates multiple tabulated equations of state that span extensive ranges of temperature, density, and composition.

Our extension of the \texttt{MESA} code is largely built upon the work of \cite{Bellinger2023}, who focused on Sun-like models containing primordial BHs. We first initialize a supermassive star with $X=0.7$, $Y=0.3$, $Z=0$. We allow the star to relax to thermal equilibrium, at which point we inject a BH of initial mass $M_\mathrm{BH,init} = 0.01\; \mathrm{M}_\star$ into the center.

\texttt{MESA} only models the portion of the star in hydrostatic equilibrium (HSE); in the case of conventional stars, the inner boundary conditions for this region, located above the inner radius $R_i$,  would be $R_i=0,M(R_i)=0,L(R_i)=0$. In the case of a quasi-star, the region in HSE forms an envelope around a non-HSE interior gravitationally dominated by the BH. As the quasi-star evolves, we calculate the BH's mass at each time step, use this to determine new inner boundary conditions for the envelope, and then have \texttt{MESA} model the envelope from those boundary conditions.

Nuclear fusion is neglected within the region dominated by the BH. In our models, hydrogen burning does occur during the relaxation phase, as well as near the base of the envelope shortly after the BH is introduced. However, temperatures in the envelope quickly drop and become too low for fusion.

We found that excessively large time steps lead to  oscillations in the envelope luminosity, density, and pressure. To mitigate this, we implemented an adaptive time-step control scheme that reduces the step size whenever the luminosity varies too rapidly. The time step is then gradually increased again until renewed fluctuations trigger another reduction. This approach effectively suppresses the variability while preserving a computationally efficient evolutionary timescale for the model.

\subsection{Inner Boundary Condition: Comparison with previous work and with new analytical results}
\label{sec:BallModel}
As a first study case, we implement the models described in \cite{Ball2011,Ball2012} (henceforth referred to as the ``Ball model"). During our work on this project, \citet{Campbell2025} released a similar implementation of \cite{Ball2011} in \texttt{MESA}. Like our implementation, theirs was based on \cite{Bellinger2023}; they additionally incorporated the Tolman-Oppenheimer-Volkoff (TOV) correction, as well as a mechanism to spatially smooth several physical quantities in order to reduce fluctuations. In light of this, we added the TOV correction to our version. We do not incorporate their smoothing mechanism, but it served as inspiration for our own solution of slowing down time steps when the fluctuations begin.

Assuming the region dominated by the BH is a sphere of some radius $R_i$, the mass at the envelope's inner boundary is
\begin{equation}
M_i=M_\mathrm{BH}+M_\mathrm{cav}\,,
\label{eq:M0}
\end{equation}
where $M_\mathrm{cav}$ corresponds to the gaseous material under the gravitational influence of the BH. Assuming $\rho(r)\propto r^{-3/2}$, this is given by $M_\mathrm{cav}=\frac{8\pi}{3}\rho_i R_i^3$, where $\rho_i$ is the density at $R_i$.

\cite{Ball2011} took $R_i$ to be a multiple of the Bondi radius \citep{Bondi1952} of the BH,
\begin{equation}
R_i = N\frac{2GM_{\mathrm{BH}}}{c_{s,i}^2},
\label{eq:bondi}
\end{equation}
where $N$ is a tunable parameter (assumed to be $N=1$ in their paper, so that $R_i$ effectively becomes the Bondi radius  of the BH). Here, $c_{s,i}$ is the sound speed at $R_i$, which is related to the density and total pressure by $c_{s,i}=\sqrt{4p_i/3\rho_i}$.

\citet{Ball2011} noted that energy produced by the BH is transmitted out of the interior via convection. Because the maximum energy flux achievable by convection is $F=pc_s$, the luminosity $L$ would be capped by the maximum convective luminosity in the region, $L_{\rm conv}=4\pi R^2_i p_ic_{s,i}$ \citep{Begelman2008}. By introducing a convective efficiency $\eta$, the accretion luminosity can be expressed as 
\begin{equation}
L=4\eta \pi R^2_i p_ic_{s,i}\,.
\label{eq:LandR}
\end{equation}
Moreover, the luminosity of the quasi-star is assumed to be produced entirely by the conversion of mass accreted by the BH into radiation. Denoting $\epsilon$ as the radiative efficiency and defining $\epsilon' = \epsilon/(1-\epsilon)$, we have $\dot{M}_\mathrm{BH} = (1-\epsilon)\dot{M}_{\rm in}$ and
\begin{equation}
L=\epsilon \dot{M}_{\rm in} c^2=\epsilon' \dot{M}_\mathrm{BH} c^2.
\label{eq:Lacc}
\end{equation}
In the above equation, $\dot{M}_{\rm in}$ represents the mass inflow rate across the base of the envelope of the quasi-star towards the BH.
Note that the system as a whole loses some mass to radiation via this mechanism, at a rate of $L/c^2=\epsilon \dot{M}_{\rm in}$.
Here, for simplicity of notation, we always indicate with $M_\star$ the total instantaneous mass of the star. 
As with \cite{Ball2011}, we assume $\epsilon = \eta = 0.1$. Using Equation \ref{eq:LandR}, we can compute $L$ once $R_i$ has been calculated (with the pressure $p_i$ and sound speed $c_{s,i}$ at the boundary being computed by \texttt{MESA}). We can then obtain $\dot{M}_{\mathrm{BH}}$ from Equation \ref{eq:Lacc}, which we use to linearly approximate the BH's mass at the next step of the model. With this, we can likewise estimate the remaining boundary conditions at the next time step.

\begin{figure}[ht!]
\includegraphics[width=\linewidth]{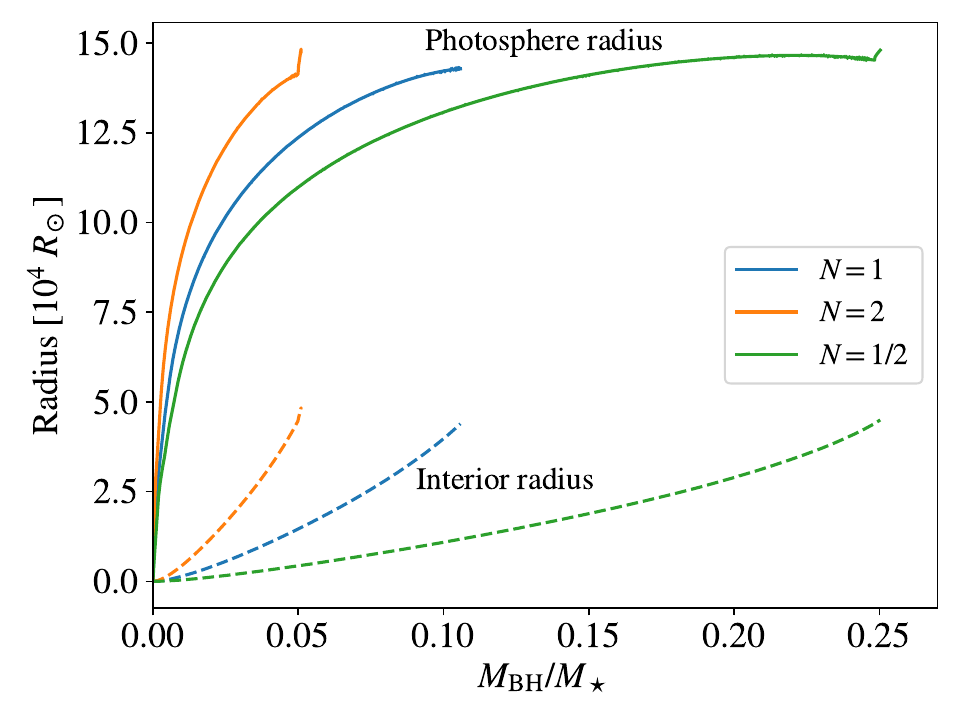}
\caption{
The radii of the photosphere (solid) and interior region $R_i$ (dashed) for a $M_\star=10^5 M_\odot$ quasi-star as a function of the BH mass normalized by the instantaneous stellar mass, with our \texttt{MESA} implementation of the \cite{Ball2011} model for different values of the parameter $N$ (cf.~Eq.~\ref{eq:bondi}). All models are initialized with the same properties and an initial BH mass of $M_{\rm BH} = 100 M_\odot$. }
\label{fig:Ball2011}
\end{figure}

Figure \ref{fig:Ball2011} shows the results for our \texttt{MESA} implementation of the model from \cite{Ball2011}, for different values of $N$ in Equation \ref{eq:bondi}. Our results are consistent with those of \cite{Ball2011} and \cite{Campbell2025}.

\cite{Ball2012} corrected the inner boundary condition of this model to account for the gravitational influence of the infalling gas, by replacing $M_\mathrm{BH}$ with $M_i=M_\mathrm{BH}+\frac{8\pi}{3}\rho_i R_i^3$ in Equation \ref{eq:bondi}. This yields a cubic equation for $R_i$, which has no physically valid solutions for BH masses above
\begin{equation}   M_{\mathrm{crit}}(N)=
\frac{c_{s,i}^3}{12\sqrt{N^3G^3 \pi \rho_i}}\,,
   \label{eq:Mcrit}
\end{equation} 
and has a unique valid solution when $M_\mathrm{BH} \leq M_{\mathrm{crit}}$:
\begin{equation}
R_i = \frac{6NGM_{\mathrm{crit}}}{c_{s,i}^2} \cos{\left ( \frac{1}{3} \arccos{\left ( \frac{M_{\mathrm{BH}}}{M_{\mathrm{crit}}} \right ) + \frac{\pi}{3}} \right )}\,.
\label{eq:r0corr}
\end{equation}
A full derivation for this is given in Appendix \ref{appendix:BondiDerivation}.

\begin{figure}[ht!]
\includegraphics[width=\linewidth]{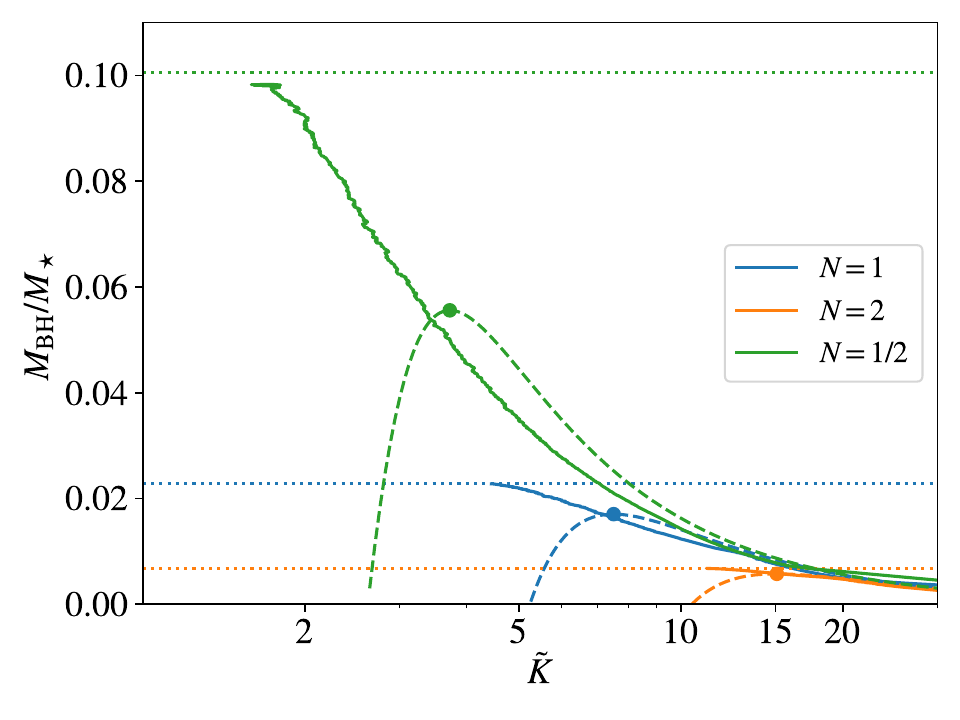}
\caption{Solid lines: Ratio between BH and star mass as a function of the dimensionless entropy, and for different values of the parameter $N$ in our \texttt{MESA} implementation of the Ball model. The numerical curves are plotted until the end of the \texttt{MESA} simulations. Dotted horizontal lines indicate our analytically derived maximum mass in the Ball model, $M_{\rm crit}(N)$ (Eq.~\ref{eq:Mcrit}). Also shown with dashed curves are the analytical solutions of \citealt{Coughlin2024}, with the maximum BH masses predicted by their solutions marked with filled circles.}
\label{fig:Ball2012}
\end{figure}

\cite{Coughlin2024} analytically estimated the maximum BH mass with this correction. They modeled the envelope with a 4/3 polytrope $p=K\rho^{4/3}$, and assumed this to be valid up to the photosphere of the quasi-star. They then computed the dimensionless entropy, defined by
\begin{equation}
\tilde{K}=\frac{4KM_i^{-2/3}}{G(4\pi)^{1/3}},
\end{equation}
and considered the maximum mass the BH could reach as $\tilde{K}$ gradually decreases. They found the maximum mass for the $N=1$ case to be on the order of $M_{\rm BH}/M_\star \sim 0.01$, approximately an order of magnitude smaller than that of the Ball~2011 model.

In Figure \ref{fig:Ball2012} we compare our \texttt{MESA} implementation of the Ball 2012 model (solid lines) with the analytical predictions of \citet{Coughlin2024} (dashed lines), as well as with our analytically derived maximum BH mass  $M_{\mathrm{crit}}(N)$ (cf.~ Eq.~\ref{eq:Mcrit}, marked with dotted lines corresponding to the various $N$). Our models were found to terminate well before reaching $R_i\approx R_\star$, with a maximum BH mass remarkably close to our analytical value $M_{\mathrm{crit}}(N)$. The final BH masses corroborate the analytical trend found by \cite{Coughlin2024} that the BHs reach higher limiting masses for lower values of $N$. However, for lower values of $N$, we reach higher BH masses than predicted by their analytical approximation. While the exact reason for this is unclear, it should be noted the analytical predictions are based on the assumptions that the convective envelope follows a $\gamma_o=4/3$ polytrope, and that the atmosphere is thin (i.e.~convective energy transfer remains efficient close to the surface of the quasi-star). We would expect any deviations from these idealizations to result in a slightly different maximum BH mass.

\subsection{Models with updated inner boundary condition}
\label{sec:CoughlinModel}
Next, we implemented the model described in \cite{Coughlin2024} (which we refer to as the ``Coughlin model"). As in the Ball model, the quasi-star is divided into an interior region dominated by a BH and an outer envelope that can be modeled by \texttt{MESA}. The luminosity is still assumed to be spatially constant and produced entirely from accretion. Physically, the difference is that the Ball model treats the interior region as a sphere of infalling gas, extending up to the Bondi radius. In contrast, the Coughlin model treats the entire interior as carrying energy at the maximum rate achievable by convection, which is expected to be more accurate for high BH masses ($M_{\rm BH}/M_\star \gtrsim 0.1$).

In terms of implementation, the Ball model requires a means to compute $R_i$ for a given BH mass (namely, the Bondi radius formula). From there, the luminosity condition is obtained by assuming $L=4\eta \pi R_i^2 p_i c_{s,i}$ at the boundary (Equation~\ref{eq:LandR}). On the other hand, the Coughlin model requires a means to compute $L$ for a given BH mass (as we discuss later, this is done by assuming $L$ equals the Eddington luminosity of the quasi-star). We can then derive $R_i$ by assuming $L=4\eta \pi r ^2 p c_s$ \textit{everywhere} in the interior region, rather than only at the boundary.

This relation between the radius, density and (spatially constant) luminosity practically serves as an equation of state for the interior region of mass $M_i$. Defining the scaled radius and mass coordinates $\xi=r/R_i$ and $m_i=M(r)/M_i$, this leads to the Lane-Emden equation for the interior:
\begin{equation}
K_i \frac{d}{d\xi}\left [\xi^{-2} \left (\frac{dm_i}{d\xi} \right )^{1/3} \right ] =-\frac{m_i}{\xi^4} \frac{dm_i}{d\xi}
\label{eq:interiorLE}
\end{equation}
where $K_i$ is a constant satisfying
\begin{equation}
R_i=M_i(GK_i)^{3/5} \left (\frac{\eta \sqrt{\gamma_i}}{L} \right)^{2/5}\,.
\label{eq:radius}
\end{equation}
Here, $\gamma_i$ is the first adiabatic index in the interior. \cite{Coughlin2024} demonstrated that $K_i$ is an invertible function of the ratio $M_{\rm BH}/M_i$.

To proceed further, assume that the convective portion of the envelope (adjacent to the interior region) can be described by a polytrope and has an adiabatic index of $\gamma_o$. Assume further that the mass, pressure and temperature are continuous when crossing $R_i$. From this, \cite{Coughlin2024} derived the Lane-Emden equation for this region:
\begin{equation}
\tilde{K}_o \frac{d}{d\xi}\left [\left (\xi^{-2}\frac{dm_o}{d\xi} \right )^{1/3} \right ] =-\frac{m_o}{\xi^2}
\label{eq:exteriorLE}
\end{equation}
where $m_o=M(r\geq R_i)/M_i$ and
\begin{equation}
\tilde{K}_o= \frac{\gamma_o}{\gamma_o-1} \frac{K_i}{m'_i(1)^{\gamma_o-\frac{1}{3}}}.
\label{eq:K_o}
\end{equation}

\cite{Coughlin2024} demonstrates that the solutions to Equations \ref{eq:interiorLE} and \ref{eq:exteriorLE} are uniquely determined by the value of $K_i$. The premise (skipping over the specifics in \citealt{Coughlin2024}) is that we can numerically solve both of these Lane-Emden equations at some $\xi=\xi_1$ in the interior and $\xi=\xi_2$ in the envelope. This gives $m_i(\xi_1)$ or $m_o(\xi_2)$, and $K_i$ can be computed from either of them.

Of course, computing $K_i$ this way would require knowledge of the values of either $m_i(\xi_1)$ or $m_o(\xi_2)$, which is not possible without knowing $M_i$ a priori. However, suppose there exists a pair of ($\xi_1$, $\xi_2$) for which we know the masses $M(\xi_1)$ and $M(\xi_2)$. $K_i$ is a function of $m_i(\xi_1)=M(\xi_1)/M_i$ and also a function of $m_i(\xi_2)=M(\xi_2)/M_i$, so it is a function of the ratio $m_i(\xi_1)/m(\xi_2)=M(\xi_1)/M(\xi_2)$, which would be calculable. Thus, we should choose a pair ($\xi_1$, $\xi_2$) for which we know the masses.

In the interior, the obvious choice is $\xi_1=0$, since $M(\xi=0)=M_{\rm BH}$ is known a priori. In the exterior, all points in the polytropic envelope satisfy
\begin{equation}
\frac{1}{\xi^{3\gamma_o-3}}\left ( \frac{m'_o(\xi)}{m'_o(1)} \right)^{\frac{3\gamma_o-1}{2}} = \frac{L_{\text{conv,max}}(r)}{L} \sqrt{\frac{\gamma_i}{\gamma_o}}\,,
\label{eq:generalstopping}
\end{equation}
where $L_{\text{conv,max}}(r)\equiv4\pi\eta r^2pc_s$ is the maximum convective luminosity achievable at radius $r$. From here, we assume $\gamma_i=\gamma_o=4/3$, which reduces the above equation to
\begin{equation}
\frac{1}{\xi}\left ( \frac{m'_o(\xi)}{m'_o(1)} \right)^{3/2} = \frac{L_{\text{conv,max}}(r)}{L}\,.
\label{eq:stopping}
\end{equation}
If we now choose $\xi_2$ to be the point where $L=L_{\text{conv,max}}$, we have
\begin{equation}
\frac{1}{\xi}\left ( \frac{m'_o(\xi)}{m'_o(1)} \right)^{3/2} = 1\,.
\label{eq:Rc}
\end{equation}

\cite{Coughlin2024} took the condition $L=L_{\text{conv,max}}$ to define the starting point of the quasi-star's radiative layer, where energy is primarily transported by radiation and the structure can no longer be described by the same polytropic model. They additionally assumed that the radiative layer was thin, such that the mass at this point is approximately $M_\star$. With these approximations, $K_i$ could be computed using $M_{\rm BH}$ and $M_\star$.

\cite{Begelman2025} performed a more careful analysis of the transition to the radiative layer, which is more correctly termed the ``inefficient convection (IC) layer'' because energy transport by convection is still significant at this radius. Denoting the thickness of the outer layer as $\Delta r=R_\star-R_{\rm IC}$, they defined the boundary between efficient convection (which can be approximated with an adiabatic equation of state) and inefficient convection to be the point at which the radiative diffusion timescale $t_{\rm diff}$ across $\Delta r$ becomes shorter than the buoyancy timescale $t_{\rm buoy}$. These timescales are defined as
\begin{equation}
t_{\rm diff} = \frac{\rho\kappa \Delta r^2}{c},\;\;\;\;\;\;
t_{\rm buoy} = R_\star \sqrt{\frac{\Delta r}{GM_\star}}.
\label{eq:timescales}
\end{equation}
Moreover, they found that about 8\% of the quasi-star's mass lies in the IC layer. This means we can no longer assume the thin atmosphere approximation ($M_{\rm IC}=M_\star$), though for the purposes of finding the interior's boundary condition, the IC layer is likely not thick enough to necessitate modeling it.

In our models, we consistently saw that the radius at which $L=L_{\rm conv,max}$ (denoted $R_c$) lies below the base of the IC layer ($R_{\rm IC}$). Since it lies within the convective envelope, though, it can still be used as a stopping point for integrating the Lane-Emden equations, provided we can compute the mass here. We opt to use $R_c$ as our stopping point rather than using $R_{\rm IC}$ (which would change the right hand side of Equation \ref{eq:Rc}), as the real envelope will likely better resemble a 4/3 polytrope here than at the base of the IC layer.

To determine the mass $M_c$ corresponding to this point, we have \texttt{MESA} find the mass coordinate at which $L=L_{\text{conv,max}}$ in each time step. From the ratio $M_{\rm BH}/M_c$, we can then numerically find $K_i$. This uniquely determines $m_{\rm BH}$, so we can then compute $M_i=M_{\rm BH}/m_{\rm BH}$. One may wonder whether this would cause issues after the interior radius $R_i$ exceeds $R_c$, since we would then not be able to use MESA to find a point in the envelope where $L=L_{\text{conv,max}}$. Fortunately, this is not a problem, because $R_c$ is in fact the theoretical limiting radius for $R_i$. This follows because $R_c$ is, by definition, the maximum radius that could possibly support saturated convection (i.e. $L=L_{\text{conv,max}}$); since the entire interior is assumed to be saturated, its boundary cannot exceed this point under the Coughlin model. As such, the Coughlin model has no theoretical solutions beyond $R_i=R_c$, corresponding to where $M_{\rm BH}/M_i=M_{\rm BH}/M_c\simeq0.62$. This ratio does not depend on the scale of the quasi-star (such as its total mass or radius), and is thus said to be a ``self-similar" property of the model across any scale.

For our implementation, we use a Mathematica notebook publicly released by Coughlin\footnote{\url{https://github.com/erc-astro/mathematica/tree/main/quasi-stars}} to perform the numerical integrations for $K_i$ and $M_i$, across a discrete set of mass ratios $M_\mathrm{BH}/M_\star$. We then perform a cubic spline interpolation on those points, allowing us to estimate these quantities at each time step.

We note that this model assumes the envelope can be approximated by a polytrope up to $R_c$, and Eq.~\ref{eq:stopping} further assumes that the exponent of the polytrope is $\gamma_o=4/3$. We considered using \texttt{MESA} to find the adiabatic index $\gamma_o$ in the envelope at each time step, and then solving the more general Eq.~\ref{eq:generalstopping} at the point where $L=L_{\rm conv,max}$. However, this would require either reintegrating Eq.~\ref{eq:generalstopping} at every time step, or solving it across a grid of $K_i$ and $\gamma_o$ and doing a 2D interpolation. To save computational speed, we use the $\gamma_o=4/3$ case at all times in the above expressions, allowing us to integrate Eq.~\ref{eq:Rc} a priori with Mathematica. For use in future work exploring a dynamic adiabatic index, we generalized the equations of the Coughlin model to arbitrary $\gamma_i$ and $\gamma_o$ in this section; they reduce to those shown in \cite{Coughlin2024} upon setting $\gamma_i=\gamma_o=4/3$.

Following \cite{Coughlin2024} and \cite{Begelman2025}, we obtain $R_i$ and $M_c$ by assuming $L$ approximately equals the Eddington luminosity of the quasi-star:
\begin{equation}
    L_{\rm Edd}=\frac{4 \pi G M_\star c}{\kappa}\,.
    \label{eq:Ledd}
\end{equation}
Here, $\kappa$ is taken at the base of the inefficient convective layer (using the definition from \citealt{Begelman2025}). Similarly to before, we use MESA to determine the point in the star at which $t_{\rm diff}=t_{\rm buoy}$, and get the opacity there. With $M_i$ and an approximation for $L$ both known, $R_i$ can then be computed via Eq.~\ref{eq:radius}.

While we assume $L=L_{\rm Edd}$ to obtain the properties of the interior, the inner luminosity boundary condition is set to be $4\eta \pi R_i^2p_ic_{s,i}$, where $p_i$ and $c_{s,i}$ are obtained from the base of the envelope using MESA. This value is generally slightly higher than the computed Eddington luminosity, but results in several properties of the interior (e.g.~$p_i$ and $c_{s,i}$) being closer to their predicted values, compared to fixing the luminosity boundary condition to $L_{\rm Edd}$. Likewise, using $L_{\rm conv,max}=L_{\rm Edd}$ to find $M_c$ is more stable and better matches predictions than using $L_{\rm conv,max}=L(r)$.

Other than the updated expressions for $M_i$, $L$ and $R_i$, and implementation of the interpolation functions, the rest of our code remains structurally the same as with the Ball model. The relation $\dot{M}_\mathrm{BH}=L/\epsilon'c^2$ is still used to obtain the BH's growth rate, with $\epsilon'$ assumed to be approximately 0.1.

\begin{figure*}[ht]
  \centering
  \begin{subfigure}[ht]{0.48\textwidth}
    \centering
    \includegraphics[width=\textwidth]{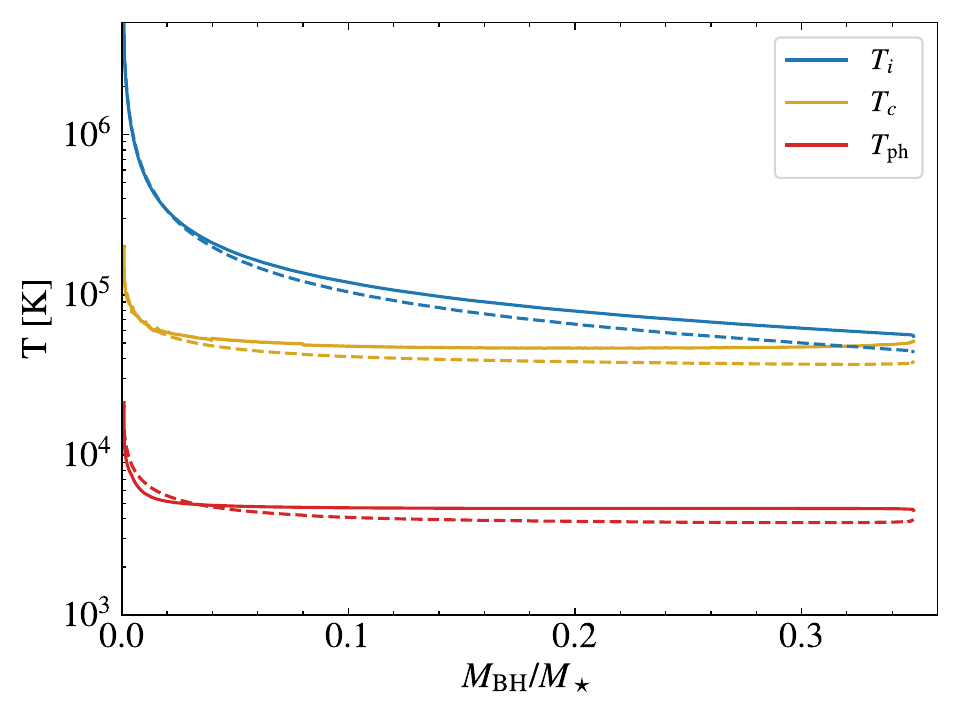}
    \label{fig:plot1}
  \end{subfigure}
  \hfill
  \begin{subfigure}[ht]{0.48\textwidth}
    \centering
    \includegraphics[width=\textwidth]{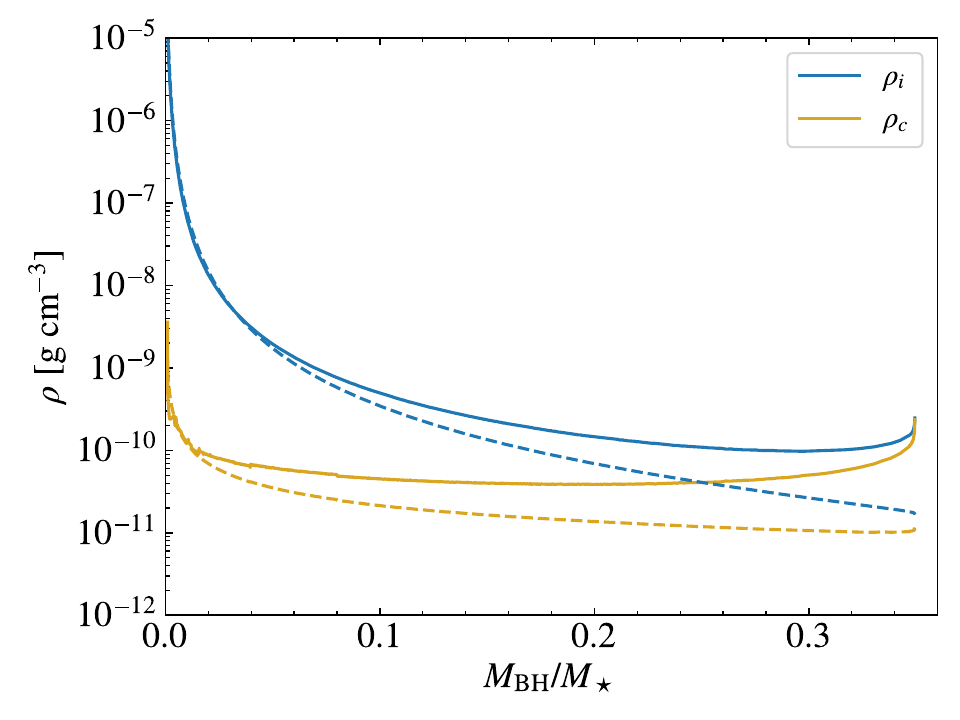}
    \label{fig:plot2}
  \end{subfigure}
  \begin{subfigure}[ht]{0.48\textwidth}
    \centering
    \includegraphics[width=\textwidth]{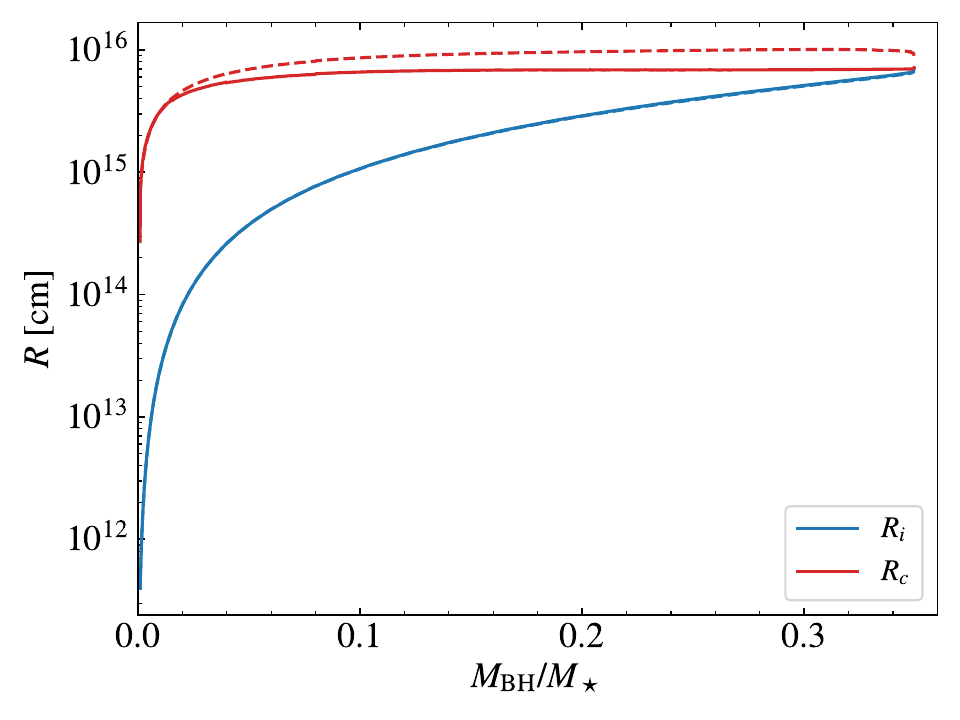}
    \label{fig:plot3}
  \end{subfigure}
  \hfill
  \begin{subfigure}[ht]{0.48\textwidth}
    \centering
    \includegraphics[width=\textwidth]{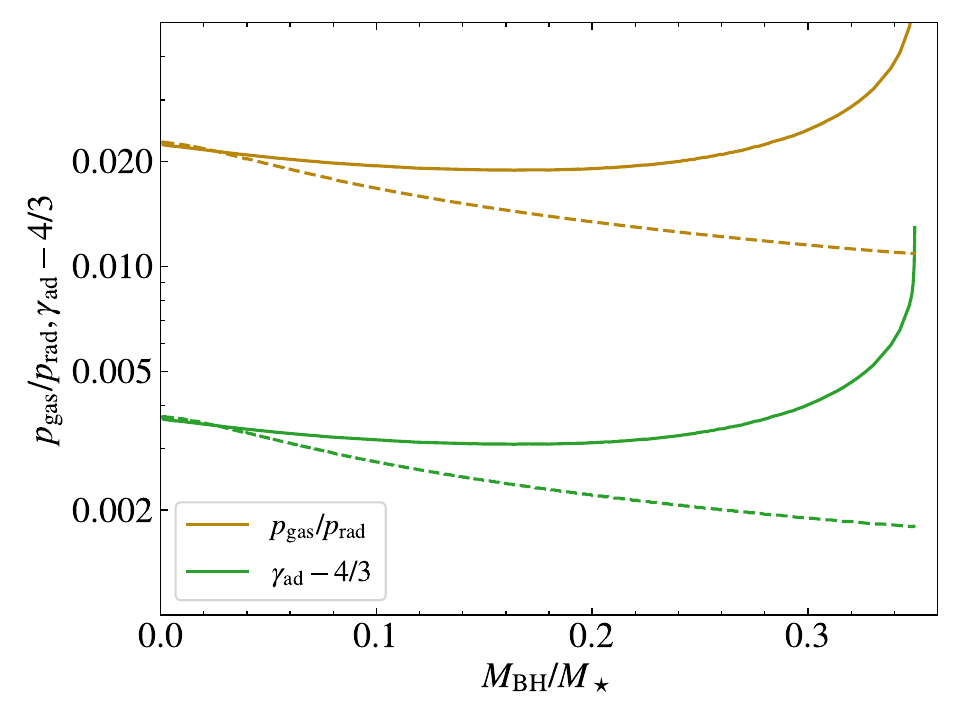}
    \label{fig:plot4}
  \end{subfigure}
  \caption{
Comparison between various quasi-star properties (from top right, clockwise: density $\rho$, pressure ratio $p_{\rm gas}/p_{\rm rad}$ and adiabatic index $\gamma_{\rm ad}$, relevant radii $R$, temperature $T$) obtained with our \texttt{MESA} implementation of the baseline model of \citet{Coughlin2024} (solid lines) and the corresponding analytical predictions from the same work (dashed lines), shown over the course of the quasi-star’s evolution. The total quasi-star mass is $M_\star=10^5M_\odot$.   
For a consistent comparison, the mass  
$M_c$ in the analytical model (i.e. the mass at which $L = L_{\rm conv,max}$; see Eq.~\ref{eq:stopping}) is gradually adjusted to track the time-dependent value from the simulation.}
  \label{fig:CoughlinResults}
\end{figure*}

\subsection{Mass Loss Implementation}
\label{sec:windstheory}
We also explore the effects of including a mass loss scheme in our implementation of the Coughlin model. Eruptive winds in massive stars might originate from regions in the stellar envelopes where the local radiative luminosity exceeds the Eddington limit, causing radiative acceleration to overcome gravity and unbind material from the star \citep{Jiang2018}.
These super-Eddington regions are usually associated with opacity peaks: in such layers, the increased opacity traps radiation, leading to density and gas-pressure inversions that destabilize the envelope \citep{paxton2013,Jiang2015,Cheng2024}. When energy transport through convection becomes inefficient (below a critical optical depth $\tau < \tau_{\rm crit}\approx c/c_s$), excess radiative flux cannot be carried by convective motions and could instead drive episodic, dynamical mass ejections.\footnote{We note, however, that this is one possible outcome. It has been pointed out that, if the mass across the outer layer is large, upwelling gas from the interior of the layer might not be propelled fast enough to escape. In this case, the radiation force imbalance might just create a turbulent but bound flow with large inhomogeneities where the radiation ultimately escapes through low-density channels with minor mass loss \citep[e.g.~][]{Dotan2012}.}
This mechanism effectively converts super-Eddington radiative luminosity into an outflow once the local radiative energy exceeds the binding energy of the overlying layers. This process is supported by 3D radiation-hydrodynamic simulations (e.g., \citealt{Jiang2015,Jiang2018}) that show how turbulence and opacity peaks can lead to outbursts and sustained envelope oscillations.
In the 1D framework implemented in \texttt{MESA} by \cite{Cheng2024}, this is modeled by identifying layers where excess radiative energy exceeds gravitational binding energy, leading to eruptive mass loss.

\begin{figure*}[ht]
  \centering
  \begin{subfigure}[ht]{0.48\textwidth}
    \centering
    \includegraphics[width=\textwidth]{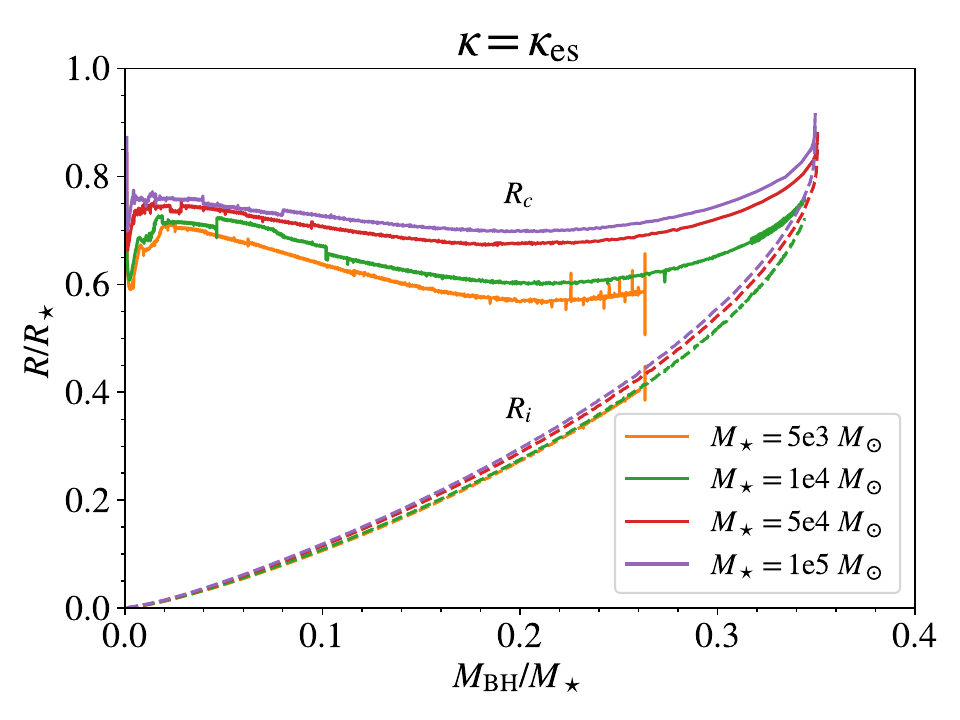}
    \label{fig:plot1}
  \end{subfigure}
  \hfill
  \begin{subfigure}[ht]{0.48\textwidth}
    \centering
    \includegraphics[width=\textwidth]{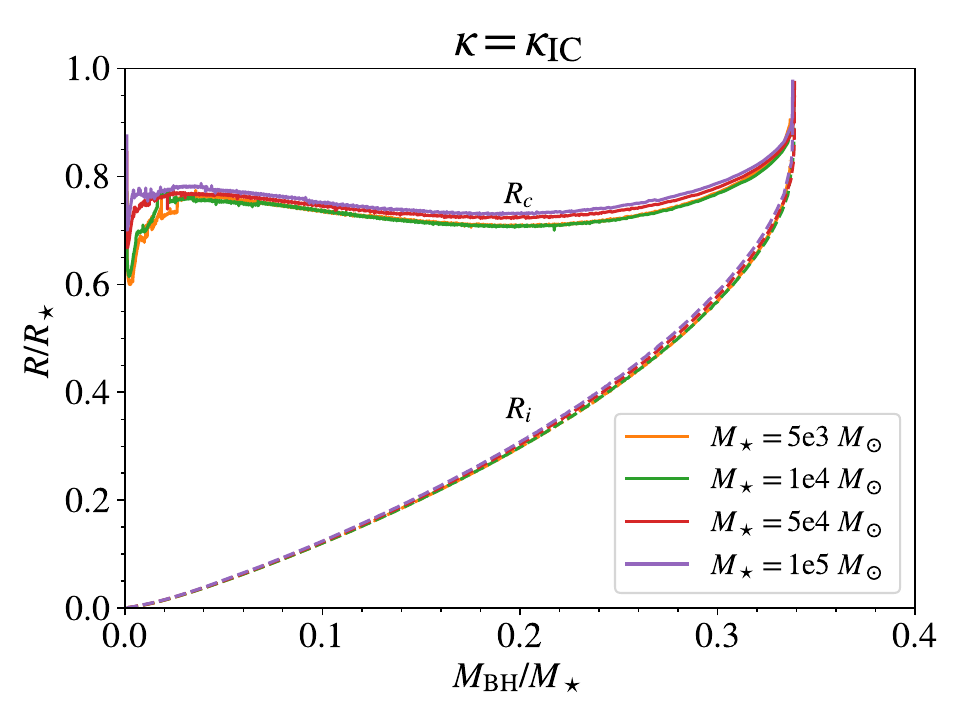}
    \label{fig:plot2}
  \end{subfigure}
  \begin{subfigure}[ht]{0.48\textwidth}
    \centering
    \includegraphics[width=\textwidth]{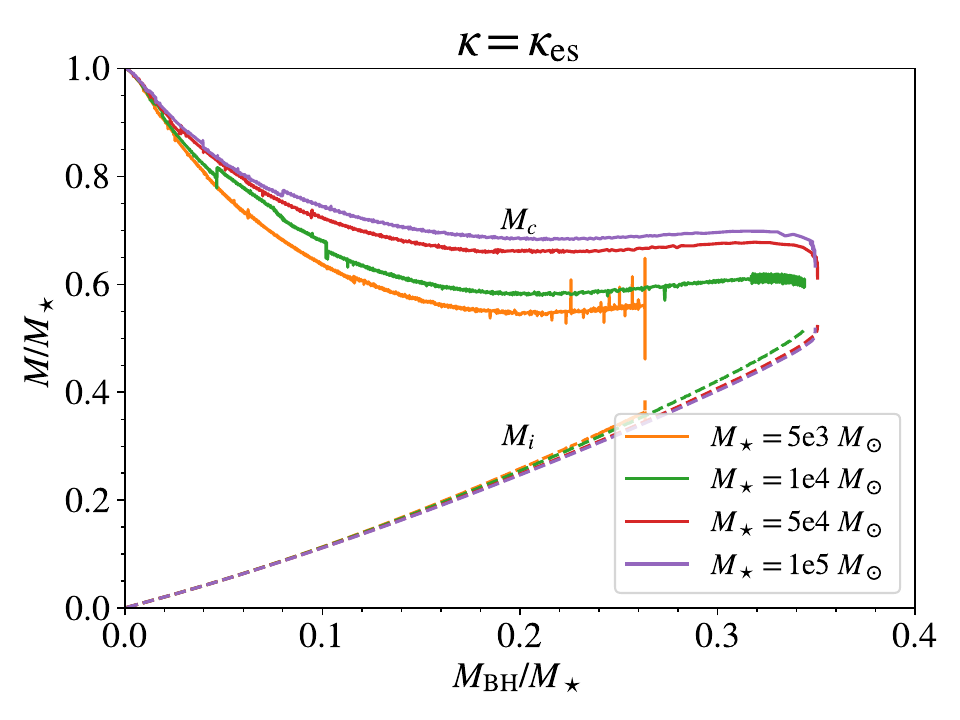}
    \label{fig:plot3}
  \end{subfigure}
  \hfill
  \begin{subfigure}[ht]{0.48\textwidth}
    \centering
    \includegraphics[width=\textwidth]{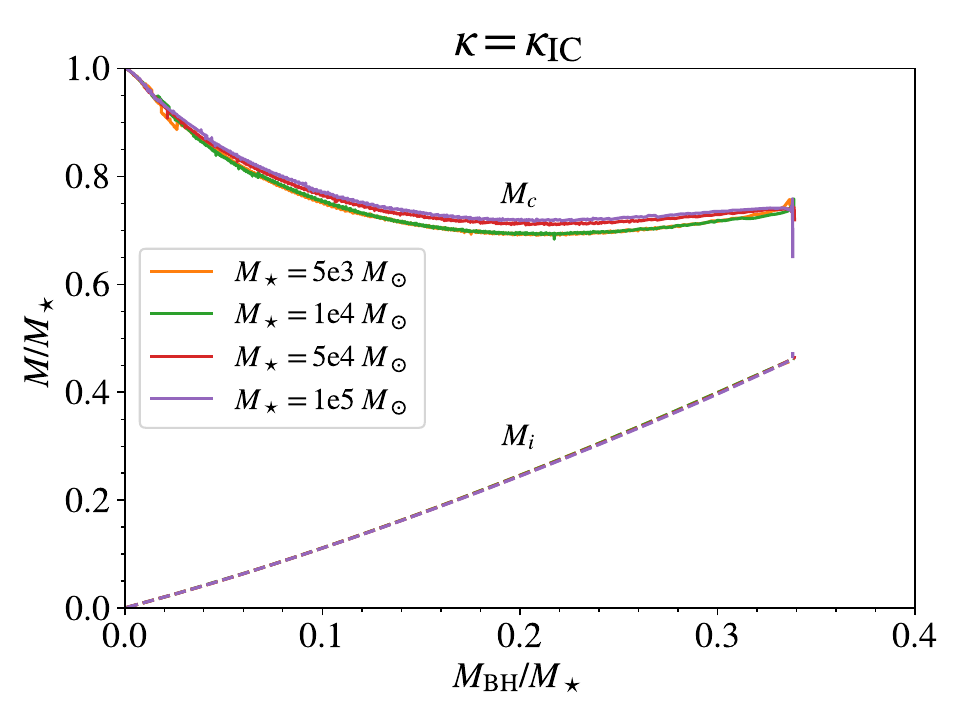}
    \label{fig:plot4}
  \end{subfigure}
  \caption{Comparison of $R_c$ with $R_i$ (top) and $M_c$ with $M_i$ (bottom), both when using $\kappa=\kappa_{\rm es}=0.34 \; \rm cm^2 \; g^{-1}$ (left) and using $\kappa=\kappa_{\rm IC}$ (right) for our calculation of $L_{\rm Edd}$, where $\kappa_{\rm IC}$ is the opacity taken at the base of the IC layer. All values shown here are normalized by the total radius or mass of the quasi-star. $R_c$ and $M_c$ are defined as the radius and mass at the point where $L_{\rm conv,max} = L_{\rm Edd}$.}
  \label{fig:convCompare}
\end{figure*}

In our implementation, we neglect the presence of other possible wind drivers and solely include this eruptive mass loss scheme using the \texttt{MESA} implementation developed by
\cite{Cheng2024}. 
A comparison of different wind prescriptions and their impact on quasi-star evolution is presented in \cite{Santarelli2025}, whose study was conducted in parallel to ours and provides a complementary reference on this topic.

\begin{figure}
    \centering
    \includegraphics[width=0.48\textwidth]{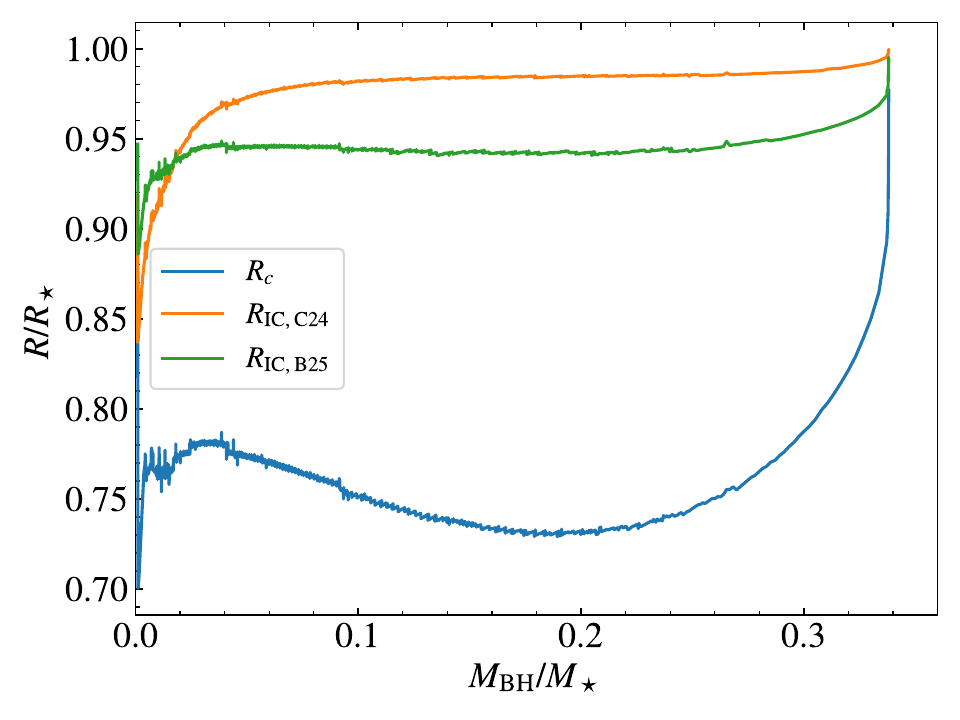}
    \includegraphics[width=0.48\textwidth]{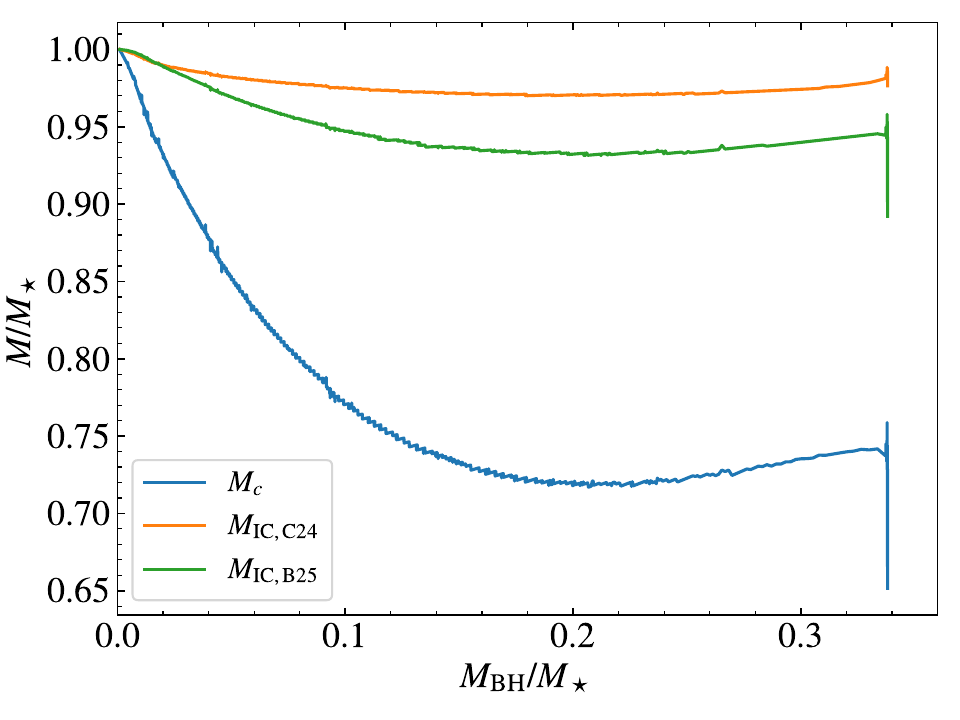}
    \caption{Top panel: Ratio of $R_c$ and $R_{\rm IC}$ to the photospheric radius, as a function of the BH mass normalized to the quasi-star total mass. We show this radius using the definitions for the IC layer from both \cite{Cheng2024} (indicated with $M_{\rm IC,C24}$) and \cite{Begelman2025} (indicated with $M_{\rm IC,B25}$). The model here is a $M_\star = 10^5 M_\odot$ quasi-star with opacity taken from the base of the IC layer in the Eddington luminosity calculation. Bottom panel: Same as above, but showing the ratio of the masses below these points to the quasi-star's total mass.}
    \label{fig:radiativeLocation}
\end{figure}

\section{Quasi-star evolution and BH growth with more realistic prescriptions}
\label{sec:results}

\subsection{Evolution with no mass loss}
\label{sec:noWinds}
As a first step, we create a model similar to that of 
\citet{Coughlin2024},
and compare our numerical evolution with \texttt{MESA}
with their semi-analytical
predictions. We use a $M_\star =10^5 M_\odot$ model and assume the electron scattering opacity $\kappa=\kappa_{\mathrm{es}}=0.34$~cm$^2$~g$^{-1}$ in Eq.~\ref{eq:Ledd}. 

\citet{Coughlin2024} assume a thin atmosphere, such that $M_c$ is practically the same as $M_\star$, and is thus kept fixed at $10^5 \; M_\odot$ for this example. As described in \citet{Begelman2025}, this approximation significantly overestimates the efficiency of convection in the star. Indeed, in our models $M_c$ is time-dependent and generally well below $M_\star$. To account for this difference, 
and thus be able to compare our numerical results with their analytical ones, 
we modify their analytical predictions to use the same $M_c$ for a given value of $M_{\rm BH}$ that we obtained from our model runs.

\begin{figure*}[ht!]
\centering
  \begin{subfigure}[ht]{0.48\textwidth}
    \centering
    \includegraphics[width=\textwidth]{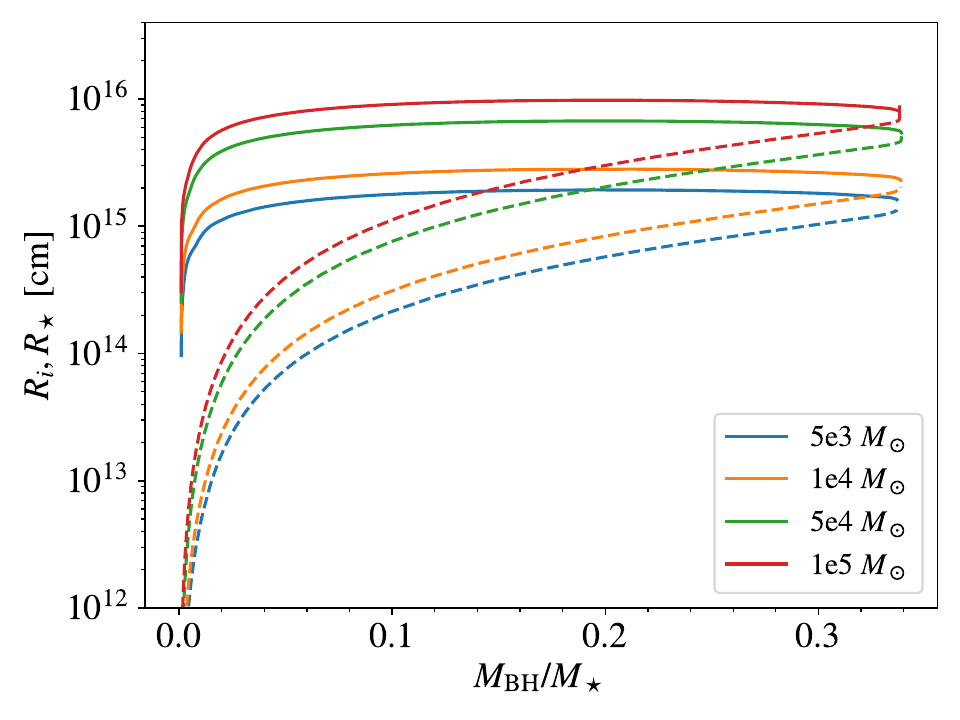}
  \end{subfigure}
  \hfill
  \begin{subfigure}[ht]{0.48\textwidth}
    \centering
    \includegraphics[width=\textwidth]{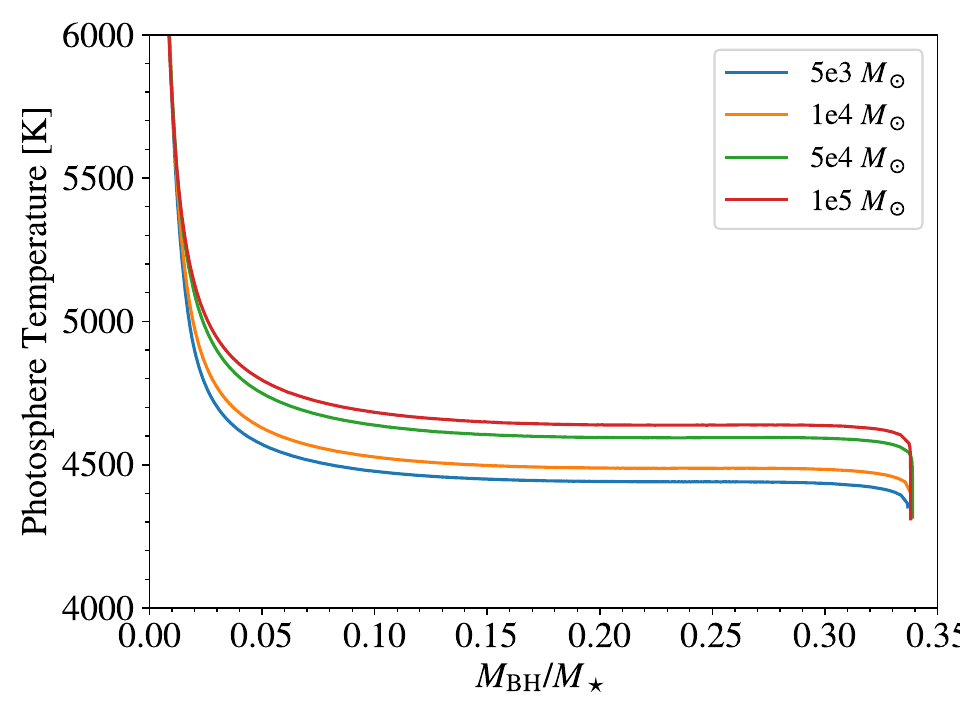}
  \end{subfigure}
\caption{Left: The radii of the photosphere $R_{\star}$ (solid) and interior region $R_i$ (dashed) for a range 
of quasi-star masses, with our \texttt{MESA} implementation of the Coughlin model. Right: photospheric temperatures for the same models.
 All models are initialized with the same properties, and an initial BH mass of $M_{\rm BH}/M_\star = 0.001$. The Eddington luminosity is calculated by taking $\kappa$ at the base of the IC layer. It is apparent that the final ratio  $M_{\rm BH}/M_\star$ is largely independent of the initial quasi-star mass.}
\label{fig:Coughlin-differentMasses}
\end{figure*}

In Figure \ref{fig:CoughlinResults}, we compare various properties of our $M_\star = 10^5 \; M_\odot$ model with the analytical predictions from Figure~6 of \cite{Coughlin2024}. As expected, our interior radii closely match the analytical predictions, since we are using the same equations for the interior as they do. On the other hand, the predicted values for $R_c$ are approximately 1.2 times larger than our results. Consequently, we reach $R_i\approx R_c$ more quickly, and our models therefore terminate at a lower ratio of $M_{\rm BH}/M_\star$ compared to the predictions in \cite{Coughlin2024} (although our final $M_{\rm BH}/M_i$ ratio of 0.66 is quite close to the predicted minimum of 0.62). Similarly, our other quasi-star properties at $R_c$ (i.e. temperature, density, pressure) also differ somewhat from the analytical predictions. 

The most noticeable of the deviations is the value for $p_{\rm gas}/p_{\rm rad}$ (and by extension, $\gamma_{\rm ad}$ as well) in the envelope, both of which are computed in the middle of the convective region. We see that our adiabatic index increases over time, as does the significance of gas pressure. Consequently, approximating the envelope with a $\gamma_o=4/3$ polytrope would be less accurate at later times. We believe this to be the cause of the deviations between our results and Coughlin. In particular, the analytical prediction defines $R_c=\xi_c R_i$, where $\xi_c$ is defined by Eq.~\ref{eq:Rc}. On the other hand, we identify $R_c$ in our numerical modeling as the radius at which $L=L_{\rm conv, max}$. While the condition $L=L_{\rm conv, max}$ is equivalent to Eq.~\ref{eq:Rc} for an ideal $\gamma_o=4/3$ polytrope as described in Section \ref{sec:CoughlinModel} (for which the mass, pressure, and density are also continuous across the interior boundary), they are not identical in \texttt{MESA}, which models the convective region using a mixing-length theory. As expected, deviations become more pronounced as $R_i$ approaches $R_c$ and convection becomes less efficient.  

Figure~\ref{fig:CoughlinResults} also shows the quasi-star properties at the interior boundary. For our models, this is obtained from the base of the envelope using \texttt{MESA}. On the other hand, the analytical predictions for the properties at the interior boundary are derived using the equation of state for the interior region. While the two are generally of the same order of magnitude, the values we obtained from just above the boundary differ somewhat from the values they calculated from just below the boundary. As such, the temperature and pressure are not perfectly continuous across the interior boundary. 

Having built a similar model to that of \citet{Coughlin2024}, and compared our results 
with the semi-analytical predictions, we next focus on more realistic models. We now account for the envelope's mass decreasing as rest energy is converted to radiation; namely, we use Eq.~\ref{eq:Lacc} to obtain $\dot{M}_\star$, and use this to linearly approximate the mass at the following time step. We also allow $\kappa$ to change over time in Eq.~\ref{eq:Ledd}, by having \texttt{MESA} sample it at the base of the IC layer (which is defined by the condition $t_{\rm diff} = t_{\rm buoy}$; see Eq.~\ref{eq:timescales}). The conversion of mass to radiation generally results in the star losing $\sim 4\%$ of its mass by the end of the simulation, which has little effect on the results. On the other hand, the use of a dynamic opacity is more significant for less massive models, as they tend to have higher absorption opacities. In our $M_\star = 10^5 \; M_\odot$ model, the opacity increases from $\sim 0.35$ to $\sim 0.50$~cm$^2$~g$^{-1}$ over time; in contrast, the opacity in our $M_\star = 10^4 \; M_\odot$ model reaches $\sim 0.75$~cm$^2$~g$^{-1}$ by its end. Because the Eddington luminosity is inversely proportional to $\kappa$, this can slightly affect $R_i$ and more noticeably affect $M_c$, as both are estimated using the Eddington luminosity.

In Figure \ref{fig:convCompare}, we compare the values of $R_c$ and $M_c$ with the radius and mass at the interior boundary, across several total quasi-star masses. We show our results when approximating $\kappa=\kappa_{\rm es}$ in our calculation for $L_{\rm Edd}$, as well as taking $\kappa$ at the base of the IC layer using \texttt{MESA}. When taking $\kappa=\kappa_{\rm es}$, we very clearly see that the region below $R_c$ is larger relative to the quasi-star in more massive models, a similar phenomenon to what occurs in sufficiently large normal stars. However, this effect is much less pronounced if we take $\kappa$ at the base of the IC layer; indeed, this increases $L_{\rm Edd}$ to a greater degree for less massive quasi-stars, and consequently increases the mass and radius below $L_{\rm conv,max} = L_{\rm Edd}$. In practice, this allows less massive quasi-stars to evolve to nearly the same $M_{\rm BH}/M_\star$ ratio as more massive stars.

We also see that $M_c$ does not quite match $M_i$ at the end of the runs. This is likely because $M_c$ is approximated here by using $L_{\rm Edd}$; while this gave us much more stable results than using $L=4\pi \eta R_i^2 \rho _i c_{s,i}$, it does overestimate the maximum possible radius that $M_i$ can reach. In all of our models, $M_{\rm BH}/M_i$ falls within the range $0.66-0.72$ at the end of the run, which is still reasonably close to the theoretical limit of $0.62$.

Because less massive quasi-stars are expected to have higher opacities, we will focus on the dynamic $\kappa$ models from here on, as we consider it the more physically motivated model.

Figure \ref{fig:radiativeLocation} shows how the mass and radius of the system below the IC layer, $M_{\rm IC}$, changes over time. For the definition from \cite{Begelman2025} (green line marked with B25), we find $M_{\rm IC}/M_\star \sim 0.93$ for most of the evolution, consistent with their estimate of the IC layer containing $\sim 8 \%$ of the quasi-star's total mass. As expected, $R_c$ (the point at which the energy transfer cannot be entirely convective) lies below the  definitions of the IC layer from both \citet{Begelman2025} and \citet{Cheng2024}, which are based on whether convective energy transfer is inefficient compared to radiative transfer. This justifies our use of Eq.~\ref{eq:Rc}, as it requires $M_c$ to lie within the convectively-efficient region. It should be noted in these plots that $R_\star$ changes over time, whereas $M_\star$ does not; for example, the photospheric radius decreases near the end of the run, while $R_c$ remains more or less static. This causes $R_c/R_\star$, and not $M_c/M_\star$, to rise to unity.

In Figure \ref{fig:Coughlin-differentMasses}, we compare the photospheric radii and temperature for models with various masses, as a function of $M_{\rm BH}/M_{\star}$. These models all use a dynamic $\kappa$; as we showed in Figure \ref{fig:convCompare}, $M_c/M_\star$ is approximately constant across the evolution in this scenario. Since the maximum $M_{\rm BH}/M_c$ is a self-similar property of the model, this results in all of these models terminating at approximately the same $M_{\rm BH}/ M_\star$ ratio.

\begin{figure*}[ht!]
\centering
  \begin{subfigure}[ht]{0.48\textwidth}
    \centering
    \includegraphics[width=\textwidth]{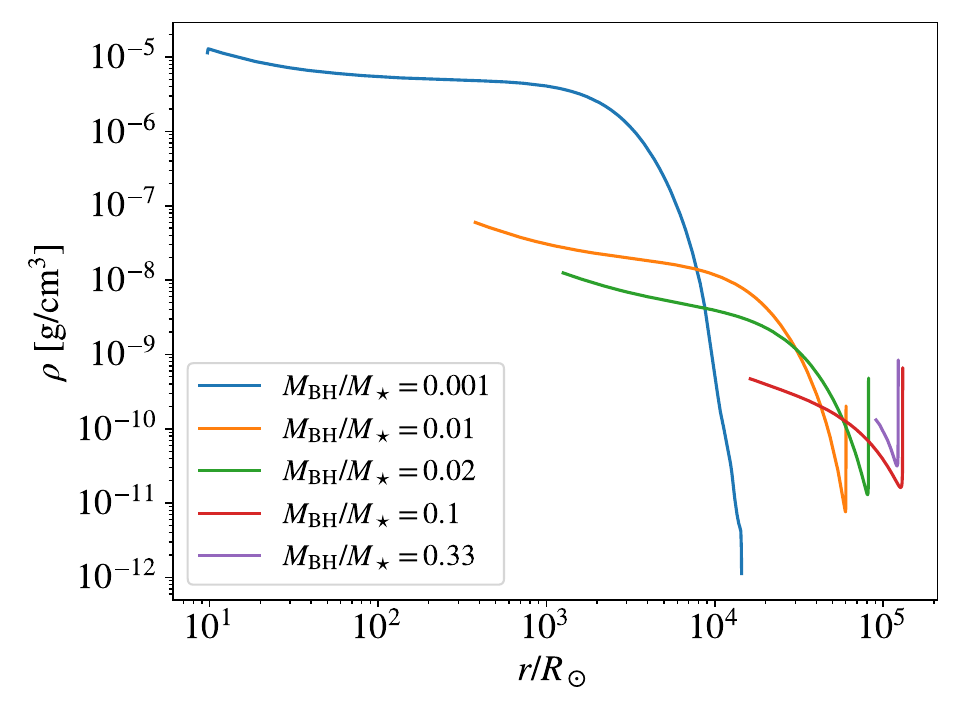}
  \end{subfigure}
  \hfill
  \begin{subfigure}[ht]{0.48\textwidth}
    \centering
    \includegraphics[width=\textwidth]{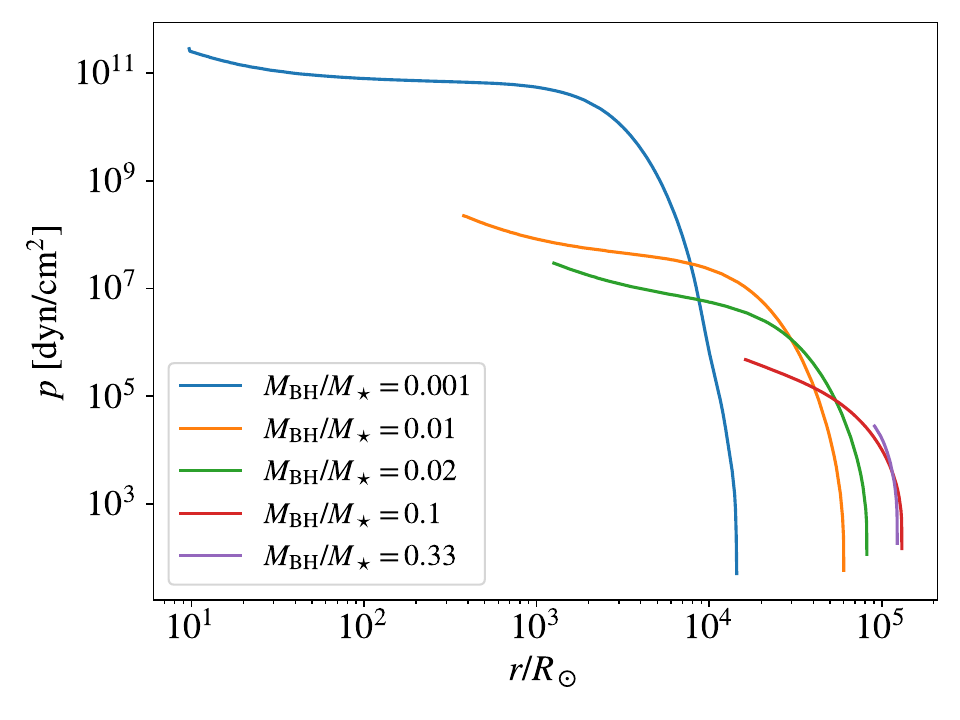}
  \end{subfigure}
\caption{Density (left) and total pressure (right) profiles for the $M_\star = 10^5 M_\odot$ Coughlin model at several values of $M_{\rm BH}/M_\star$ during the quasi-star evolution. Only the density at radii $r>R_i$ is plotted.
Note the density inversion which starts to appear during the later phases of evolution when convection becomes inefficient near the surface. }
\label{fig:Coughlin-profiles}
\end{figure*}

In Figure \ref{fig:Coughlin-profiles}, we show density and pressure profiles for our $M_\star =10^5 M_\odot$ model. While the density profile is virtually monotonically decreasing for very small BH masses ($M_{\rm BH}/M_\star \sim 10^{-3}$), density inversions begin to manifest near the surface for $M_{\rm BH}/M_\star \sim 10^{-2}$ onwards. As described in \cite{Begelman2025} and \cite{Cheng2024}, the existence of these inversions is expected near the transition to the IC layer: convection becomes inefficient near the surface, and the radiative flux slightly exceeds the local Eddington value. The resulting imbalance leads to a region where the outward radiation force drives material upward, and 
hydrostatic equilibrium can only be maintained if the gas pressure increases outward, implying a density inversion (a rise in density with radius).
 On the other hand, the profile for the total pressure does not display such an inversion, as a result of the fact that \texttt{MESA}
 strives to maintain hydrostatic equilibrium.

\subsection{Evolution with mass loss}
\label{sec:winds}
\begin{figure}
    \centering
    \includegraphics[width=0.48\textwidth]{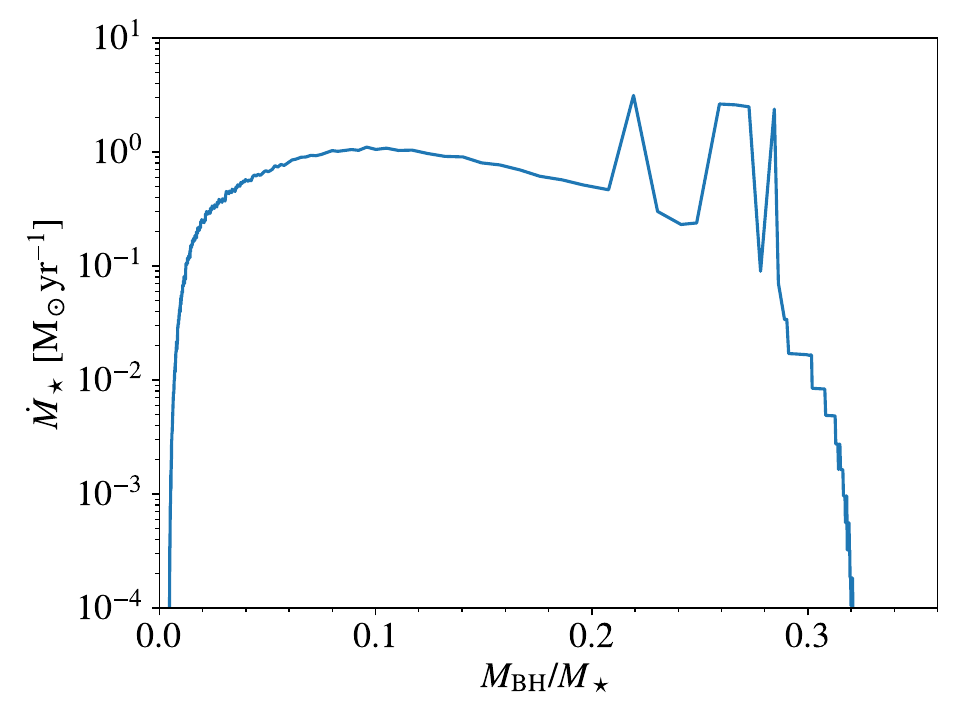}
    \caption{Wind mass loss rate during the quasi-star evolution, for a Coughlin model with initial mass $M_\star=10^5 M_\odot$ and mass loss due to eruptive winds.}
    \label{fig:mdotWinds}
\end{figure}

\begin{figure*}[ht!]
    \centering
    \includegraphics[width=\textwidth]{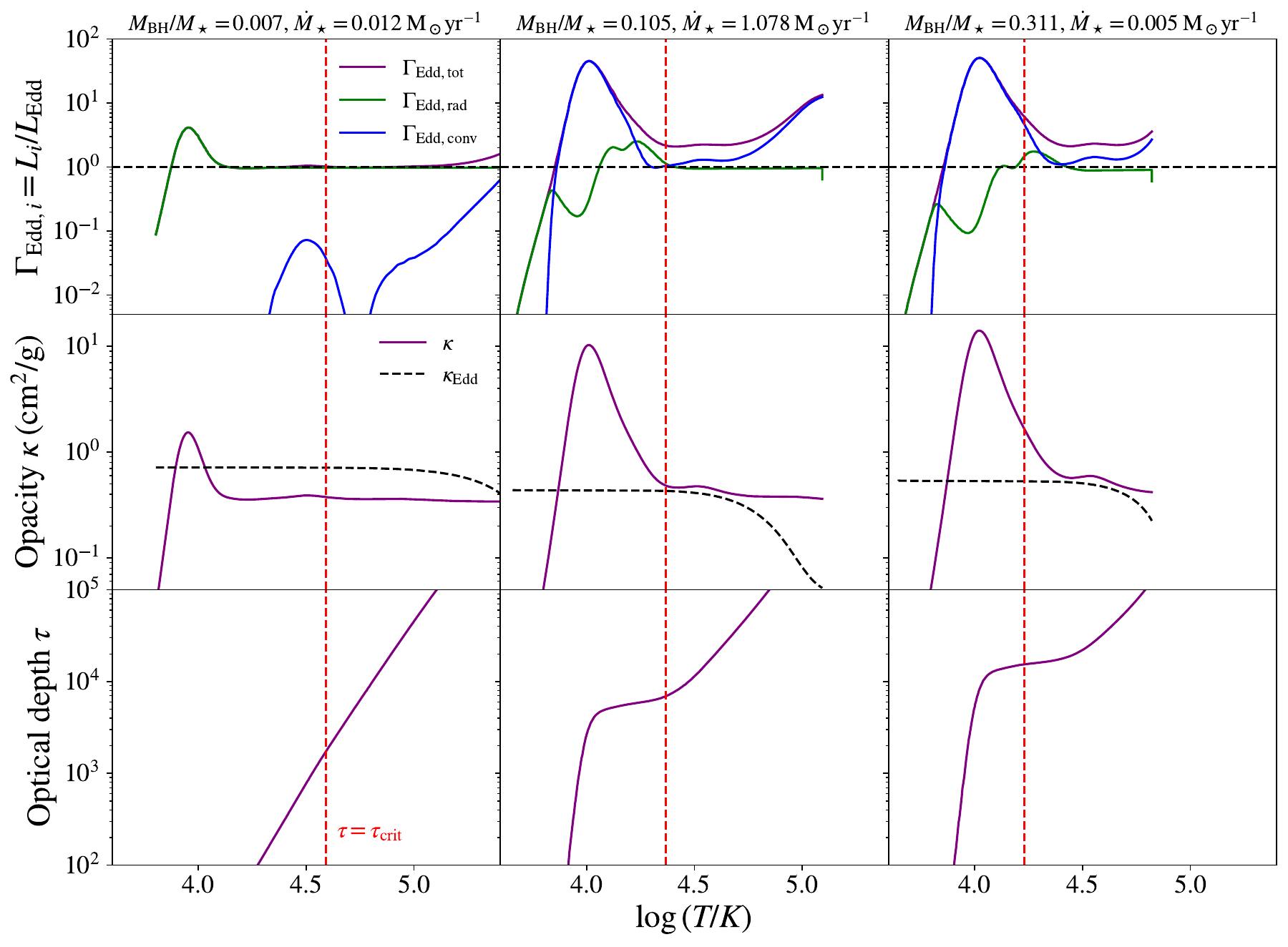}
    \caption{Profiles of luminosity in Eddington units (top row), opacity (middle row), and optical depth (bottom row) for our $10^5 \; M_\odot$ quasi-star model with eruptive winds enabled. Left column: Snapshot taken during the start of wind mass loss ($M_{\rm BH}/M_\star = 0.007$); Middle column: Snapshot taken during the presence of strong winds ($M_{\rm BH}/M_\star = 0.105$); Right column: Snapshot when winds have almost entirely ceased ($M_{\rm BH}/M_\star = 0.311$). The vertical red dashed line shows the location of the convectively-inefficient layer ($\tau < \tau_{crit}$). Eruptive mass loss only occurs if $\Gamma_{\rm Edd, rad}$ exceeds 1 in this part of the model, 
    and ceases entirely when  $\Gamma_{\rm Edd,rad}\leq1$ to the left of the red line .}
    \label{fig:EddingtonFactor}
\end{figure*}

\begin{figure}
    \centering
    \includegraphics[width=0.48\textwidth]{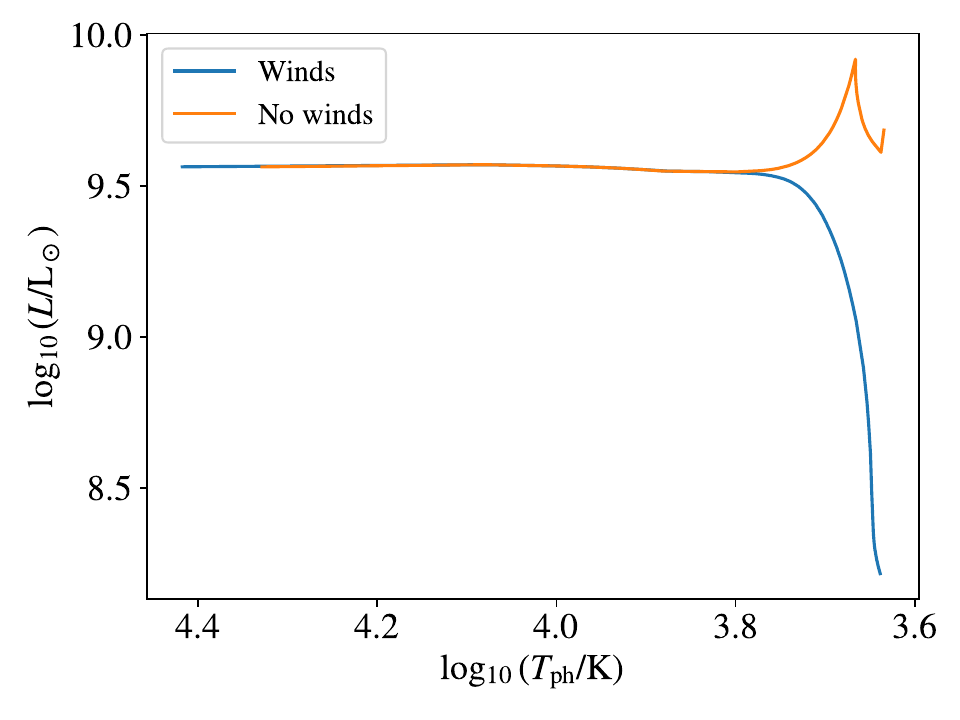}
    \caption{Hertzsprung–Russell diagram of a quasi-star of initial mass $10^5 \; M_\odot$, evolving according the the Coughlin model, and considering both models with and without mass loss.}
    \label{fig:HRdiagram}
\end{figure}

For our $M_\star =10^5 \; M_\odot$ Coughlin model, we also implemented the wind scheme from \cite{Cheng2024}  described in Section \ref{sec:windstheory}. We take the wind efficiency factor to be 0.1 (see Eq.~12 of \citealt{Cheng2024}).

The predicted mass loss rate as a function of 
$M_{\rm BH}/M_\star$ during the quasi-star evolution is shown in Figure~\ref{fig:mdotWinds}. Winds manifest early on, and cease at approximately $M_{\rm BH}/M_\star \sim 0.31$. We find that these winds greatly reduce the quasi-star's mass over time, leaving it with a total mass of $\sim 4.5\times 10^3 M_\odot$ by the end of the run (less than 1/20 of its original mass). 

Figure \ref{fig:EddingtonFactor} compares the Eddington factor $\Gamma=L/L_{\rm Edd}$ throughout the quasi-star with the temperature profile, both during the onset, peak, and end of the wind mass loss. The start of mass loss corresponds to the formation of a spike in opacity, causing $\Gamma_{\rm Edd, rad}$ to rise in the IC layer. As the quasi-star evolves, the IC layer becomes increasingly dense, causing the optical depths in this region to rise. $\tau_{\rm crit}$ also increases, albeit at a slower rate than the optical depths near the surface of the quasi-star; this causes the region for which $\tau<\tau_{\rm crit}$ (in other words, the IC layer) to shrink. Eventually, the IC layer becomes so thin that $\Gamma_{\rm Edd, rad}<1$ everywhere in it, causing the winds to cease. At this point, we still have $\Gamma_{\rm Edd, rad}>1$ slightly below the IC layer, but super-Eddington luminosities below the IC layer are not assumed to contribute to winds in this model.

\begin{figure}
    \centering
    \includegraphics[width=0.48\textwidth]{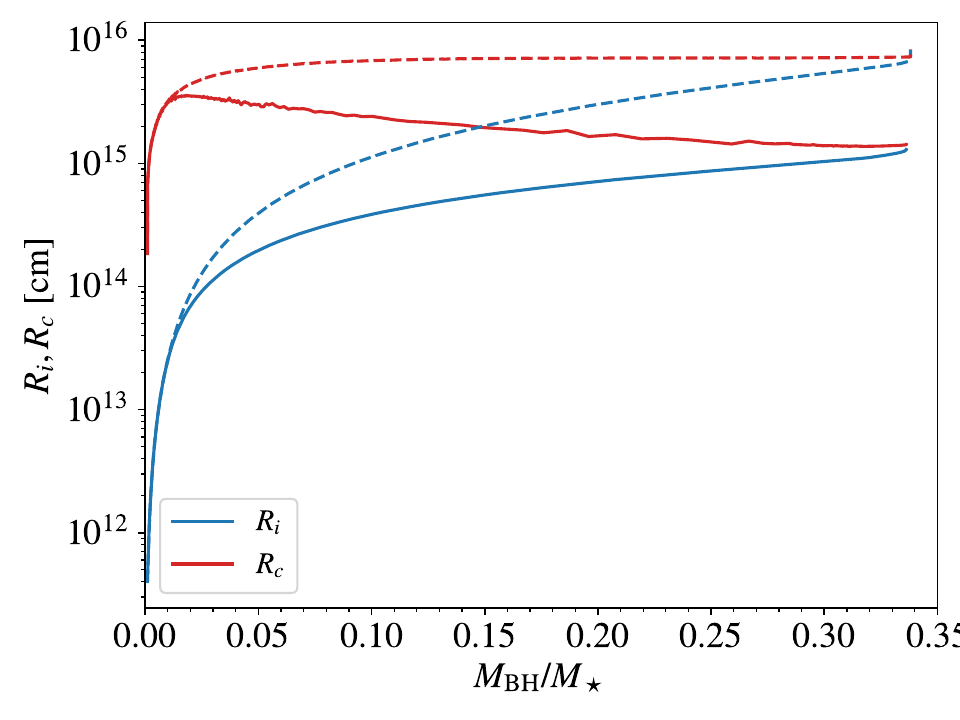}
    \caption{Values of $R_c$ and $R_i$ for the Coughlin model with winds (solid lines) and without winds (dashed lines), as a function of the ratio of the BH mass to the quasi-star mass.
    The initial quasi-star mass 
    is $M_\star=10^5 M_\odot$.
    In the case with winds, this decreases with time. }
    \label{fig:radiusWinds}
\end{figure}

Figure \ref{fig:HRdiagram} shows the evolution of the quasi-star's luminosity versus its photospheric temperature (effectively an ``HR diagram'') for the same $M_\star =10^5 \; M_\odot$ model, with and without winds. Once the BH is injected, the star gradually expands and progresses rightward toward cooler effective temperatures, with its luminosity approximately at the Eddington limit. At an age of $t\approx 1.5\times 10^5 \; \rm yr$, the opacity sharply spikes up near the surface of the quasi-star and a density inversion forms. In the model without winds, we see this correspond to an increase in luminosity past the Eddington limit. Near the end of the quasi-star's lifetime, the luminosity then begins to drop, corresponding to a drop in
temperature (cfr. Figure~\ref{fig:Coughlin-differentMasses}, right panel).
With the implementation of stellar winds, on the other hand, the opacity spike triggers eruptive mass loss. As the mass $M_\star$ drops, so does the Eddington limit and therefore the luminosity.

\begin{figure*}[ht]
  \centering
  \begin{subfigure}[ht]{0.48\textwidth}
    \centering
    \includegraphics[width=\textwidth]{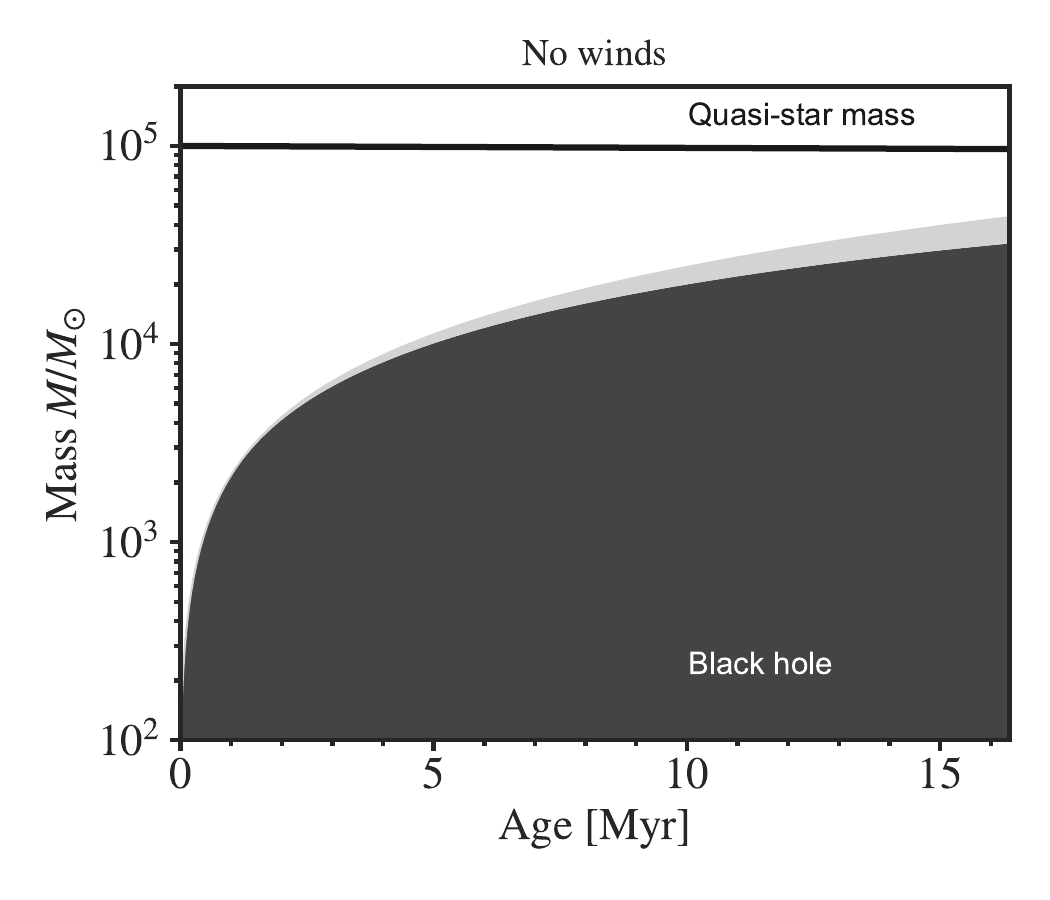}
    \label{fig:plot1}
  \end{subfigure}
  \hfill
  \begin{subfigure}[ht]{0.48\textwidth}
    \centering
    \includegraphics[width=\textwidth]{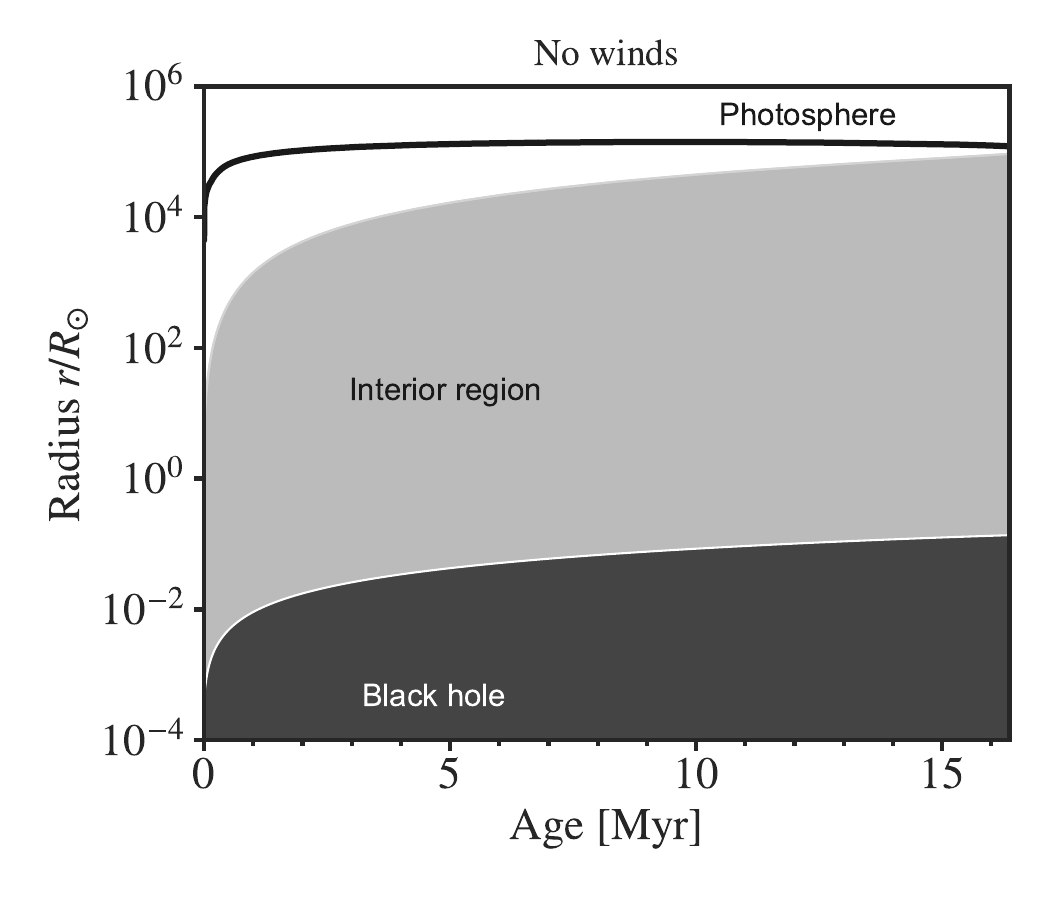}
    \label{fig:plot2}
  \end{subfigure}
  \begin{subfigure}[ht]{0.48\textwidth}
    \centering
    \includegraphics[width=\textwidth]{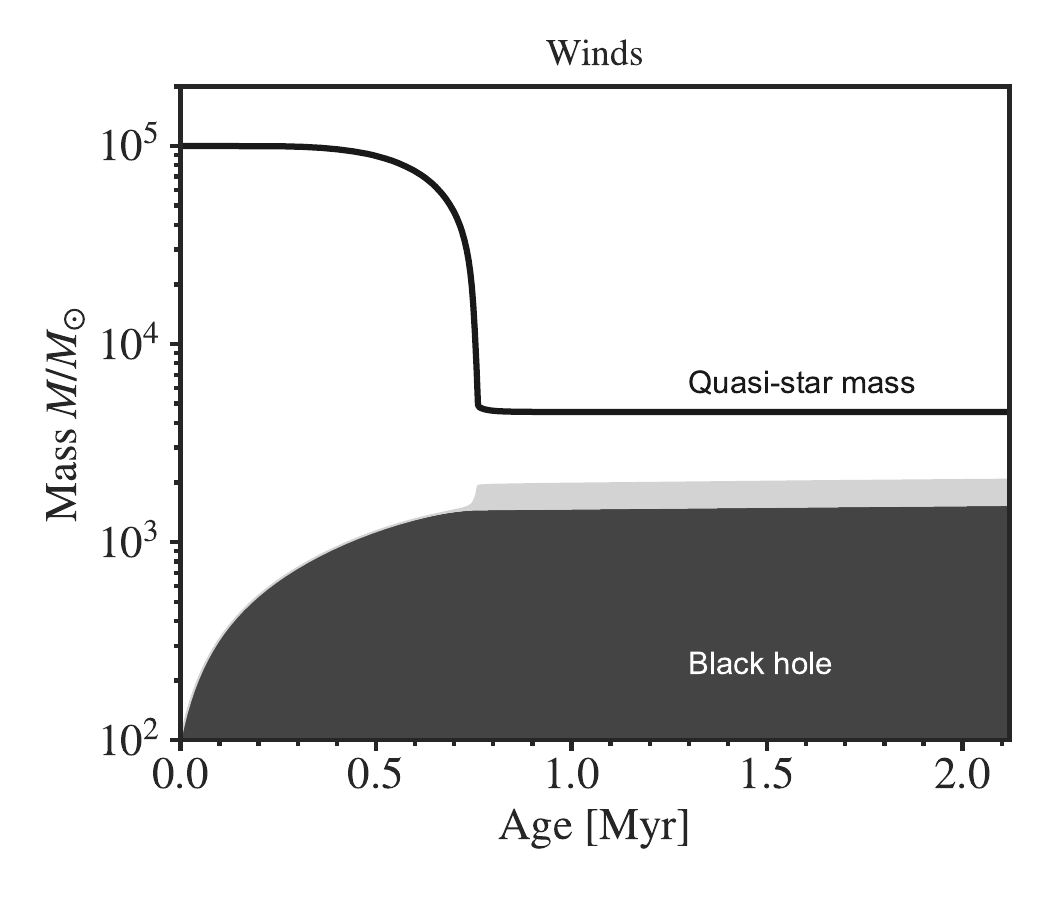}
    \label{fig:plot3}
  \end{subfigure}
  \hfill
  \begin{subfigure}[ht]{0.48\textwidth}
    \centering
    \includegraphics[width=\textwidth]{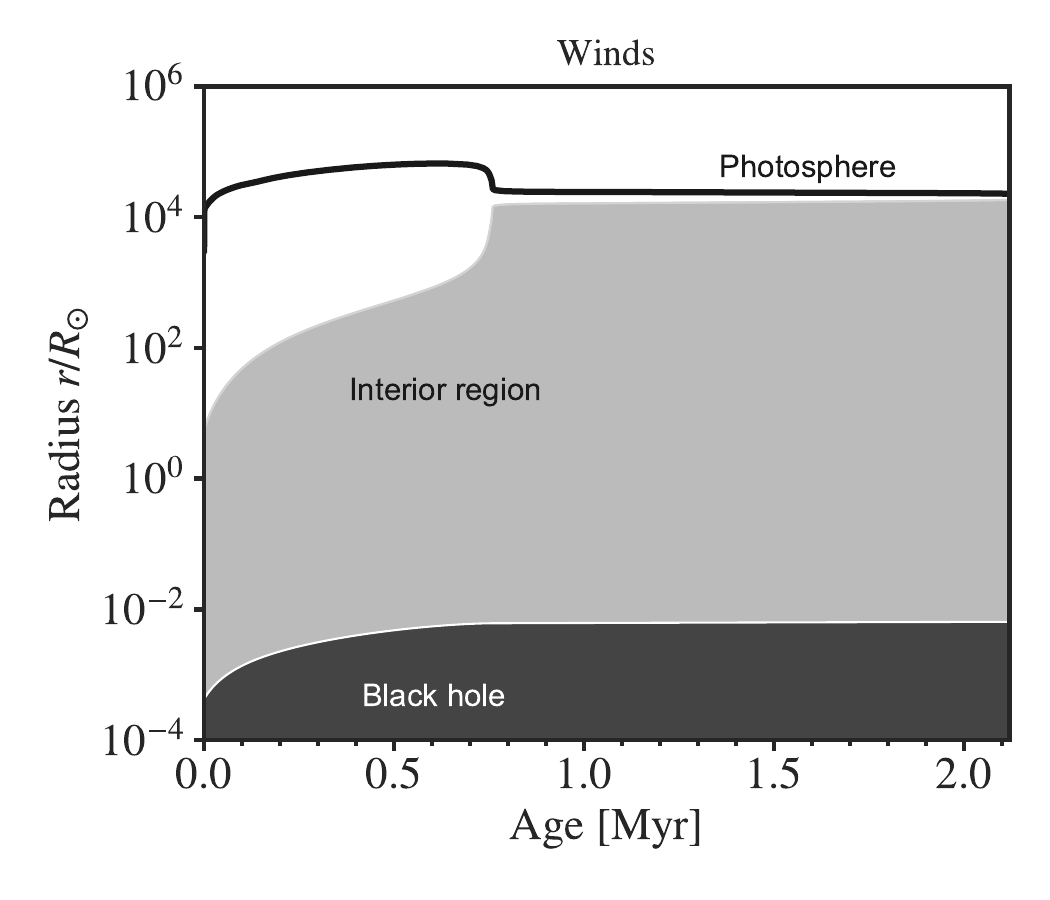}
    \label{fig:plot4}
  \end{subfigure} 
  \caption{Kippenhahn diagrams for the $M_\star = 10^5 \; M_\odot$ model, both without winds (top) and with winds (bottom). Dark gray regions correspond to the BH; light gray corresponds to the saturated convection region, and white corresponds to the envelope.}
  \label{fig:Kippenhahn}
\end{figure*}

Figure \ref{fig:radiusWinds} shows the evolution of $R_c$ and $R_i$ for the same $M_{\star,\rm init} = 10^5 M_\odot$ model, with winds (solid) and without winds (dashed). Winds begin to cause significant deviations from the fixed $M_\star$ model starting around $M_{\rm BH}/M_\star \sim 0.01$. Near the end of the evolution, at around $M_{\rm BH}/M_\star \sim 0.31$, mass loss quickly drops, causing both $R_i$ and $R_c$ to suddenly start evolving as they would for a fixed (albeit now much smaller) $M_\star$. It is important to point out that, although $M_{\rm BH}/M_\star$ is used as the horizontal axis, $M_\star$ is virtually constant for the non-wind model but is changing for the wind model. Thus, the wind model actually has a significantly shorter lifespan than the non-wind model (2.1 Myr for the wind model, compared to 16.6 Myr for the non-wind model).

As noted earlier, certain aspects of the Coughlin solution are self-similar, meaning they are consistent across different physical scales (e.g. for different values of $M_\star$). For example, the final value of $M_{\rm BH}/M_i$ is largely independent of the value of $M_\star$. Since $\kappa$ increases over time at the base of the IC layer, this model also ends at a very similar (albeit slightly smaller) value of $M_{\rm BH}/M_c$ to the non-winds model (cf.~Fig~\ref{fig:convCompare}).

Figure \ref{fig:Kippenhahn} shows Kippenhahn diagrams for the $10^5 M_\odot$ quasi-star model, both with and without winds. These diagrams
display how different regions inside a star evolve as the star ages.
 Prior to the onset of mass loss, the BH and the saturated convection region gradually grow over time. In the non-wind model, we see that the BH initially comprises almost all of the interior's mass, and gas becomes an increasingly significant portion of it over time.

Once mass loss begins, $M_\star$ decreases rapidly, causing a concomitant reduction in the Eddington luminosity of the quasi-star
 (cf.~Eq.~\ref{eq:Ledd}).
The growth rate of the BH likewise drops, as it is proportional to the luminosity (Eq.~\ref{eq:Lacc}). While the mass of the BH remains nearly constant during this brief period, the mass of the quasi-star (and consequently, $M_c$ as well) greatly decreases, and the ratio $M_{\rm BH}/M_c$ therefore increases. In other words, our model quickly shifts from being a large quasi-star with a relatively small BH, to being a smaller quasi-star with a relatively large BH. After the winds finally cease, the BH continues evolving as it would for a $M_\star=4.5\times 10^3  M_\odot$ quasi-star late in its life.

\section{Summary and future directions}
\label{sec:summary}
The discovery of extremely massive BHs at very high redshifts has posed a long-standing problem regarding how such objects could have formed within a short time after the beginning of the universe.
 This issue has become more pressing in recent years with the JWST discovery of the LRDs, for which rapidly accreting BHs represent a compelling interpretation.
Among the proposed scenarios of early BH growth, quasi-stars \citep{Begelman2008, Begelman2025} constitute a prominent candidate.

In this work, we have presented a numerical investigation of quasi-stars through dedicated modifications to the public stellar evolution code \texttt{MESA}. We examined the growth of the central BH under various assumptions for the inner boundary condition, determined by the physical properties of the region surrounding the accreting BH. In addition, we investigated the influence of the outer convectively-inefficient layer and the potential impact of mass loss through winds.

Our investigation proceeded through increasing levels of complexity, and our main findings are summarized as follows: 
\begin{itemize}
    \item We began by implementing the quasi-star models by \cite{Ball2011,Ball2012}, characterized by an inner boundary condition set at a multiple $N$ of the Bondi radius. For a fixed total quasi-star mass, our numerical simulations for the \cite{Ball2012} model consistently terminated when the black hole mass reached a critical value that depends on $N$, given by $M_{\mathrm{crit}}(N)=c_{s,i}^3/(12\sqrt{N^3G^3\pi\rho_i})$, which is a limit we independently derived analytically. Although the dependence of the maximum mass on $N$ 
    has previously been noted and derived by \citet{Coughlin2024}, our analytical expression provides a closer match to the numerical results, particularly for smaller values of $N$. 
    \item We next implemented in \texttt{MESA} a quasi-star model developed by \cite{Coughlin2024}, in which 
an inner convective region is matched to an outer adiabatic envelope. We found that the BH grows to a fraction $\approx 33\%$ of the quasi-star mass, roughly independent of the initial mass of the quasi-star. 
We demonstrated that the factor of $\sim 2$ discrepancy between our maximum BH mass and that derived by \citet{Coughlin2024} arises from their assumption that $M_c$, the maximum mass at which convection could account for all the energy transfer in the envelope, remains equal to $M_\star$ throughout the quasi-star evolution. In other words, they assume $M_i$ can grow up to $M_\star$, so that their predicted limit of $M_{\rm BH}/M_i \sim 0.62$ is equivalent to $M_{\rm BH}/M_\star \sim 0.62$. We do not make this assumption, so while our final $M_{\rm BH}/M_i$ is close to the prediction, our final $M_{\rm BH}/M_\star$ differs from it.

We also sought to better estimate the star's luminosity in the Coughlin model, which is approximated by the Eddington limit. Instead of assuming $\kappa=\kappa_{\rm es}=0.34 \; \rm cm^2 \; g^{-1}$, we have \texttt{MESA} find the opacity at the base of the IC layer. When making this adjustment, our models terminated at roughly $M_{\rm BH}/M_\star \sim 0.33$ across all our choices of $M_\star$.

The photospheric temperatures of quasi-stars are found to decrease over time and, except during the very early stages of their evolution, lie in the range 
$\sim 4500-4800$~K for the range of quasi-star masses
we studied, with higher masses naturally corresponding to higher effective temperatures.

    \item 
We next implemented eruptive winds into our \texttt{MESA} version of the Coughlin quasi-star model, following the prescription of \citet{Cheng2024}, and examined their impact for a fiducial case with $M_\star=10^5 M_\odot$.
We find that the inclusion of winds not only shortens the quasi-star's lifetime but can also lead to substantial mass depletion. In the case studied, the final stellar mass was reduced to approximately one-twentieth of its initial value. Notably, the final BH–to–quasi-star mass ratio remains roughly the same as in the wind-free case, as do other self-similar properties of the Coughlin model.
            
\end{itemize}

Our investigation represents an additional step toward understanding black hole growth within quasi-stars, and thereby their potential role as progenitors of supermassive BHs in the early universe, as well as possible sources of LRDs,
should these indeed correspond to accreting BHs as suggested by some observational evidence.

We anticipate several directions for future work. First, since quasi-stars are expected to form in very dense environments, accretion onto them is likely to play a significant role. In fact, the presence of accretion during the quasi-star's lifetime appears to be essential for substantial BH growth, given the potentially disruptive effect of winds, which can greatly reduce the quasi-star’s mass and, consequently, the mass of the BH forming within it.

Another important step toward connecting quasi-stars to their observable properties, and thus more quantitatively assessing the suggestion by \citet{Begelman2025} that they may be progenitors of LRDs, will be a more detailed treatment of their outer layers, and particularly the opacity, with the aim of predicting the spectral energy distributions of quasi-stars.

The overall stability of quasi-stars presents another intriguing avenue for investigation. Hyper-accretion onto BHs is known to drive powerful outflows and jets. If the infalling gas possesses angular momentum, an accretion disk may form around the central BH, creating conditions conducive to jet formation. Whether a quasi-star can remain stable over an extended period under such circumstances—and, if so, what its resulting observational signatures might be—remains an open question for future study.

\begin{acknowledgments}
JBH and RP acknowledge support from NASA award 80NSSC25K7554.
JBH and PJA acknowledge support from award 644616 from the Simons Foundation. MCB acknowledges support from NASA Astrophysics Theory Program grants 80NSSC22K0826 and 80NSSC24K0940.
The Center for Computational Astrophysics at the Flatiron Institute is supported by the Simons Foundation.
\end{acknowledgments}

\bibliography{biblio}

\appendix

\section{Derivation of corrected Bondi radius}
\label{appendix:BondiDerivation}
In \citet{Ball2012}, the Bondi radius is given by
$$R_i = N \frac{2G(M_{\rm{BH}}+M_{\rm{cav}})}{c_{s,i}^2}
    = N \frac{2G}{c_{s,i}^2} \left( M_{\rm{BH}} + \frac{8\pi \rho_i}{3} R_i^3 \right)$$
$$R_i^3 - \frac{3c_{s,i}^2}{16 \pi N G \rho_i} R_i + \frac{3M_{\text{BH}}}{8\pi \rho_i} = 0$$
The discriminant of this depressed cubic is
$$\Delta = \frac{243}{64 \pi^2 \rho_i^2} \left( M_{\text{crit}}^2 - M_{\text{BH}}^2 \right)$$
where we define $M_{\mathrm{crit}}=c_{s,i}^3/12\sqrt{N^3G^3 \pi \rho_i}$.

Consider the case $M_{\text{BH}} > M_{\text{crit}}$. Then $\Delta < 0$, so by Cardano’s formula,
$$R_i = \left( -A - \sqrt{A^2 - B^3} \right)^{1/3} 
     + \left( -A + \sqrt{A^2 - B^3} \right)^{1/3}$$
where we denote
$$A = \frac{3 M_{\text{BH}}}{16 \pi \rho_i}
\quad
B = \frac{c_{s,i}^2}{16 \pi \rho_i NG}.$$

Since $R_i$ must be positive, it follows that
$$\left( -A - \sqrt{A^2 - B^3} \right)^{1/3} 
    > - \left( -A + \sqrt{A^2 - B^3} \right)^{1/3}$$
$$- A - \sqrt{A^2 - B^3} > A - \sqrt{A^2 - B^3}$$
$$- A > A$$
This means $A$ must be negative, which is unphysical (since $A$ is defined in terms of positive quantities). Thus, there is no valid solution for $M_{\text{BH}} > M_{\text{crit}}$.

Now consider the case $M_{\text{BH}} = M_{\text{crit}}$. Then $\Delta = 0$, so the roots of the cubic are
$$R_i = \frac{3NGM_{\text{crit}}}{c_{s,i}^2}, 
\quad -\frac{6NGM_{\text{crit}}}{c_{s,i}^2}.$$
Obviously, only the positive solution can be physical.

Finally, consider the case $M_{\text{BH}} < M_{\text{crit}}$. Then $\Delta > 0$, so there are three distinct trigonometric solutions. For $k=0,1,2$, we have
$$R_i = \frac{6NGM_{\text{crit}}}{c_{s,i}^2} 
\cos \left( \frac{\arccos(-M_{\text{BH}}/M_{\text{crit}}) - 2\pi k}{3} \right)$$
$$= \frac{6NGM_{\text{crit}}}{c_{s,i}^2} 
\cos \left( \frac{\arccos(M_{\text{BH}}/M_{\text{crit}}) + \pi (2k-1)}{3} \right).$$
The only choice satisfying the boundary condition $R_i(M_{\text{BH}} \to 0) = 0$ is $k=1$. Thus, this is the only physically valid trigonometric solution:
$$R_i = \frac{6NGM_{\text{crit}}}{c_{s,i}^2} 
\cos \left( \frac{\arccos(M_{\text{BH}}/M_{\text{crit}}) + \pi}{3} \right).$$

If we set $M_{\text{BH}} = M_{\text{crit}}$ in this formula, we recover $R_i = 3NGM_{\text{crit}}/{c_{s,i}^2}$. Hence, the Bondi radius is given by this formula for $M_{\text{BH}} \leq M_{\text{crit}}$, and there are no physical solutions for $M_{\text{BH}} > M_{\text{crit}}$.

\end{document}